\documentclass[conference]{IEEEtran}

\makeatletter
\def\ps@headings{%
\def\@oddhead{\mbox{}\scriptsize\rightmark \hfil \thepage}%
\def\@evenhead{\scriptsize\thepage \hfil \leftmark\mbox{}}%
\def\@oddfoot{}%
\def\@evenfoot{}}
\makeatother
\usepackage{booktabs} %
\usepackage{booktabs} %
\usepackage{graphicx}
\usepackage{hhline}
\usepackage{pbox}
\usepackage{rotating}
\usepackage{array,graphicx}
\usepackage{multirow}
\usepackage{tabularx}
\usepackage{algpseudocode}

\usepackage{hyperref}
\usepackage{graphicx}
\usepackage{subcaption}
\usepackage{underscore}
\usepackage{wrapfig}
\usepackage{amsmath,amssymb,amsfonts}
\usepackage{algorithm}
\usepackage{graphicx}
\usepackage{textcomp}
\usepackage{xcolor}

\usepackage[top=0.7in, bottom=1in, left=0.625in, right=0.625in]{geometry}

\usepackage{caption}
\usepackage[backend=bibtex, style=ieee, urldate=comp, natbib=true]{biblatex}
\def\BibTeX{{\rm B\kern-.05em{\sc i\kern-.025em b}\kern-.08em
    T\kern-.1667em\lower.7ex\hbox{E}\kern-.125emX}}
\begin{document}

\title{Learning the Associations of MITRE ATT\&CK Adversarial Techniques\\
\thanks{Identify applicable funding agency here. If none, delete this.}
}

\author{\IEEEauthorblockN{Rawan Al-Shaer}
\IEEEauthorblockA{\textit{Electrical \& Computer Engineering} \\
\textit{Carnegie Mellon University}\\
Pittsburgh, United States \\
reshaer@andrew.cmu.edu}
\and
\IEEEauthorblockN{Jonathan M. Spring}
\IEEEauthorblockA{\textit{CERT/CC, SEI} \\
\textit{Carnegie Mellon University}\\
Pittsburgh, United States \\
jspring@sei.cmu.edu}
\and
\IEEEauthorblockN{Eliana Christou}
\IEEEauthorblockA{\textit{Department of Mathematics \& Statistics} \\
\textit{University of North Carolina Charlotte}\\
Charlotte, United States \\
echris15@uncc.edu}

}

\maketitle
\vspace{-2em}

\begin{abstract}
The MITRE ATT\&CK Framework provides a rich and actionable repository of adversarial tactics, techniques, and procedures (TTP). However, this information would be highly useful for attack diagnosis (i.e., forensics) and mitigation (i.e., intrusion response) if we can reliably construct technique associations that will enable predicting unobserved attack techniques based on observed ones. 
In this paper, we present our statistical machine learning analysis on APT and Software attack data reported by MITRE ATT\&CK to infer the technique clustering that represents the significant correlation that can be used for technique prediction. Due to the complex multidimensional relationships between techniques, many of the traditional clustering methods could not obtain usable associations. Our approach, using  hierarchical clustering for inferring attack technique associations with $95\%$ confidence, provides statistically significant and explainable technique correlations. Our analysis discovers $98$ different technique associations (i.e., clusters) for both APT and Software attacks. Our evaluation results show that $78\%$ of the techniques associated by our algorithm exhibit significant mutual information that indicates reasonably high predictability.

\end{abstract}
\maketitle

%
\section{Introduction}
As cyber attacks increase in volume and sophistication, the state of the art of cybersecurity solutions is still lagging behind. According to Red Canary, Advanced Persistent Threat (APT) attacks have increased from approximately 500 attacks per year in 2009 to almost 1500 APT attacks per year in 2016. Verizon and SANS reported that in 2017 about 68\% of attacks went undiscovered, and even after detection, about 77\% of these attacks took hours to days to get remediated. The lack of timely detection and response is mainly caused by the insufficient support of attack action correlations and prediction to allow for proactive intrusion investigation and mitigation. 

MITRE has created a public knowledge-based repository of adversary tactics and techniques called MITRE ATT\&CK. As of February 2020, MITRE ATT\&CK provides a total of 440 attack ``techniques" belonging to 27 different ``tactics". A tactic is a behavior that supports a strategic goal; a technique is a possible method of executing a tactic.  Each technique has a description explaining what the technique is, how it may be executed, when it may be used, and various ``procedures" for performing it. There are 174 techniques belonging to 15 pre-attack tactics, and 266 techniques belonging to 12 post-exploit (Enterprise) tactics~\cite{mitre}. For example, the {\em PRE-T1345 (Create Custom Payloads)} pre-attack technique, which belongs to the {\em Build Capabilities} tactic, describes how an adversary creates custom payloads with which malware can perform malicious actions. Also, the post-exploit technique {\em T1003 (Credential Dumping)}, which belongs to the {\em Credential Access} tactic, describes how an adversary obtains account login and password information in the form of a hash or clear text password from the operating system and software. A sequence of techniques from different tactics used for an attack is what we call a TTP (Tactics, Techniques, Procedures) chain. Moreover, the combination of MITRE ATT\&CK techniques in a TTP chain represents various attack scenarios that can be composed in an attack graph~\cite{spring2018litrev}.

MITRE ATT\&CK techniques and procedures provide behavioral observables for detecting attacks by analyzing cyber artifacts collected from the network and end-system. The structure of TTP allows analysts to organize which adversarial actions belong to specific procedures that relate to specific techniques and tactics, and helps the analyst to understand what an adversary may be trying to achieve and how to better defend against it. 
MITRE ATT\&CK highlights many techniques an adversary might use, but does not provide sufficient tips on how the adversary might combine different techniques to accomplish their goals. Thus, the technique \textit{associations} an analyst needs to construct TTP chains remain underspecified. Technique associations are important because they help an analyst to reason about adversarial behavior and predict unobserved techniques based on the observed ones in the TTP Chain. Without technique associations, an analyst will struggle to reason efficiently about adversarial behavior as the search space grows too large as the number of TTP chains increases exponentially with the number of given techniques. 
Since little to no work has been done regarding technique correlations, this paper focuses on learning the attack technique associations that manifest technique inter-dependencies and relationships based on data of real-life attacks (66 Advanced Persistent Threats, and 204 attacks by Software).

For these reasons, we developed a novel approach using hierarchical clustering to infer technique associations that represent various technique inter-dependencies in a TTP chain.
This paper's key contribution is discovering, with $95\%$ confidence, the fine-grain technique associations that support predictability of attack behavior.  
Moreover, the hierarchical clustering developed in this work can also be used to infer the coarse-grain associations between techniques across different clusters. The fine-grain associations demonstrate sequential, disjunctive, or conjunctive relationships; the coarse-grain associations demonstrate complementary relationships that complete the attack chain.

To address these goals, we first explored various partitioned clustering methods in order to group attack techniques that are likely to co-occur together in the TTP Chain. We analyzed a dataset of 270 APT and Software attacks which include a total of 345 techniques using K-means clustering~\cite{kmeans}, Partition Around Medoids (PAM)~\cite{pam}, and Fuzzy clustering~\cite{fuzzy}. However, we found that, due to the high dimensional data, such clustering techniques could not reveal practically meaningful associations, as the clusters are highly overlapping and difficult to distinguish. We then developed hierarchical clustering, which is more likely to be a suitable approach to represent the sophisticated attack patterns, especially in APT attacks, since it can model the correlations at various levels relative to the height of the hierarchy.

There are three key challenges to achieve accurate research results given our goals. First, we address the importance of establishing a suitable distance metric for clustering in terms of technique correlations in attack TTP chains while maintaining interpretability. We integrate existing distance metrics (Jaccard~\cite{jaccard} distance and Phi Coefficient~\cite{phi} correlational distance) that measure the co-occurrences and co-absences of the techniques into clustering algorithms. 
Second, we address the multi-dimensional relationships exhibited by attack techniques by extending the agglomerative hierarchical clustering method to obtain accurate results. In particular, we explore a range of hierarchical linkage approaches to establish the optimal method for binary multi-dimensional data.  As a result, our developed approach obtains the most compact technique associations.
Third, we address the challenge of validating the stability and significance of the learned hierarchical clustering and technique associations using a statistical hypothesis test. 
Our goal is to ensure that the resulting associations will be statistically significant at a rigorous confidence level ensuring that no more than 5\% of the results are due to random chance. 

This paper is organized as follows: Section~\ref{Datasets} introduces the datasets used in this work, along with their challenges and limitations. 
Section \ref{sec:complixity} presents the preliminary analysis and Section \ref{Inferring Attack Technique Associations} presents our approach to inferring technique associations. Sections \ref{Evaluation} and \ref{discussion} present the evaluation and the findings of our approach, respectively.

\section{Datasets}
\label{Datasets}
We analyzed two different datasets from MITRE ATT\&CK. The datasets referenced 270 total attack instances, made up of 209 unique techniques. In this section, we will describe the nature of the datasets in more detail, as well as the challenges they pose and their limitations.

\subsection {Description}
\label{Dataset Description}
 Stemming from MITRE's classification of reported attacks, the categories of the instances in the datasets are Advanced Persistent Threats (APTs) and Software attacks. APT attacks are synonymous to MITRE's terminology of threat actor ``Groups", while Software attacks encompass malware, ransomware, trojans, Remote Access Tools (RATs), and other code used for malicious purposes. 

Each dataset consists of all the post-exploit techniques which compose the TTP chain of any given APT or Software. In the datasets, we treat each TTP chain as an attack instance and every technique as a feature. The datasets are composed of discrete variables, specifically, asymmetrical binary variables. The outcome of the features is either 0 or 1, representing the negative or positive occurrence of a technique in an attack instance, respectively. An asymmetrical binary dataset refers to a structure such that an outcome of 1, or the positive occurrence of a technique, is more informative than an outcome of 0.

The first dataset contains 66 APT attack instances as published in the MITRE ATT\&CK Framework~\cite{mitre} on June 30, 2019. These APT attacks were mapped by MITRE from publicly reported technique use, where the original references are included in each technique description. 
The second dataset contains 204 Software attack instances as published in the MITRE ATT\&CK Framework ~\cite{mitre} on July 30, 2019. MITRE also mapped Software attacks from publicly reported technique use and accounts for the capability of the software adversary to use a technique. We downloaded the datasets from MITRE ATT\&CK in json format using the ATT\&CK Navigator~\cite{navigator}.

\subsection{Challenges}
\label{Dataset Challenges}
We faced two challenges with the datasets. The first challenge involved adjusting for the differing complexities among the Software and APT attack instances. The second challenge addressed MITRE's constant updates to the ATT\&CK framework.

For the first challenge, we addressed the ramifications of less composite attack instances. Since APT attack instances represent a full campaign containing several high-level and high-impact goals, their TTP chain is more complex than that of Software attack instances, which are typically carried out for specific low-level purposes. This implicit difference between APT and Software attacks is reflected where many of the Software attack instances are comprised of very few techniques. Attack instances with scarce technique occurrences can alter our results and create misleading technique associations due to less complex Software attacks reporting the use of techniques that are not combined in any meaningful way. For that reason, we decided to only include Software attack instances which employ at least five different techniques across five different tactics.

For the second challenge, we addressed the common updates to the ATT\&CK Framework and the discovery of newly reported attacks. We emphasize a transparent analysis methodology, as recommended by \cite{stodden2015reproducing}, so our work could be repeated as the datasets evolve.

~\subsection{Limitations}
The datasets have two salient limitations. The first limitation is described by MITRE as a limit of their data collection process. MITRE states that the APT and Software attacks are not representative of all possible techniques used by the actors associated with the observed data, rather a subset of what has been available through public and open source reporting. Therefore, the actual techniques these attacks utilized, or the ground truth, is difficult to determine or even discover. 

The second limitation of the attacks in the datasets is mapping biases. We recognize that heuristics and automated mappings of threat reports to techniques may inadvertently possess a degree of proclivity. 
For these reasons, our analysis provides an approach for characterizing APT and Software attacks that can constantly be enhanced. %

\section{Preliminary Analysis}
\label{sec:complixity}
Our process of knowledge discovery through clustering involved several stages of analysis. In this section, we will discuss our preliminary analysis using partitioned clustering, particularly K-means~\cite{kmeans}, PAM~\cite{pam}, and Fuzzy~\cite{fuzzy} clustering. The preliminary results show the true complexity of inferring accurate and applicable technique associations.

\subsection{Partitioned Clustering of Techniques}
\label{partitioned}

Partitioned clustering is used to classify observations into groups, or clusters, based on their similarity. The three partitioned clustering methods we investigated require choosing the optimal number of clusters, $K$, as described below in Section \ref{optimalK}. The details of the clustering algorithms will be discussed in Section \ref{Partitioned Algorithms}, and the validation methods are discussed in Section \ref{Clustering Validation}.

\subsubsection{{Choosing the Optimal K}}
\label{optimalK}
In order to perform any type of partitioned clustering, the number of clusters must be specified in advance. While there are numerous ways to determine the optimal number of clusters, we found the Elbow and Silhouette methods appropriate for the nature of the datasets. Throughout this paper, we will use $k$ for any number of clusters, and $K$ for the optimal number of clusters in partitioned clustering.

The Elbow method chooses $K$ by minimizing the within-cluster variance, which is calculated using Euclidean distances, a distance metric inappropriate for binary data; see Remark \ref{Unsuitable Distance Metrics} for details. Thus, we customized the use of the Elbow method by incorporating the dissimilarity matrix. Specifically, we computed the cluster variance by summing the squared dissimilarity divided by the size of each cluster. Then, as represented by the name, $K$ is determined by the elbow, the point at which the cluster variance by dissimilarity suddenly decreases. 

The Silhouette method measures how well a data point belongs in a cluster by computing the pairwise distances between all the objects in every cluster, using any specified distance metric. In order to achieve well clustered results, the distance between the clusters, known as the silhouette width, should be maximized. For that reason, $K$ is chosen as the point with the highest silhouette width. 

\begin{figure}[t]
     \centering
     \begin{subfigure}[b]{0.23\textwidth}
         \centering
         \includegraphics[width=\textwidth, height=3cm, trim={0 0 0 12cm}, clip]{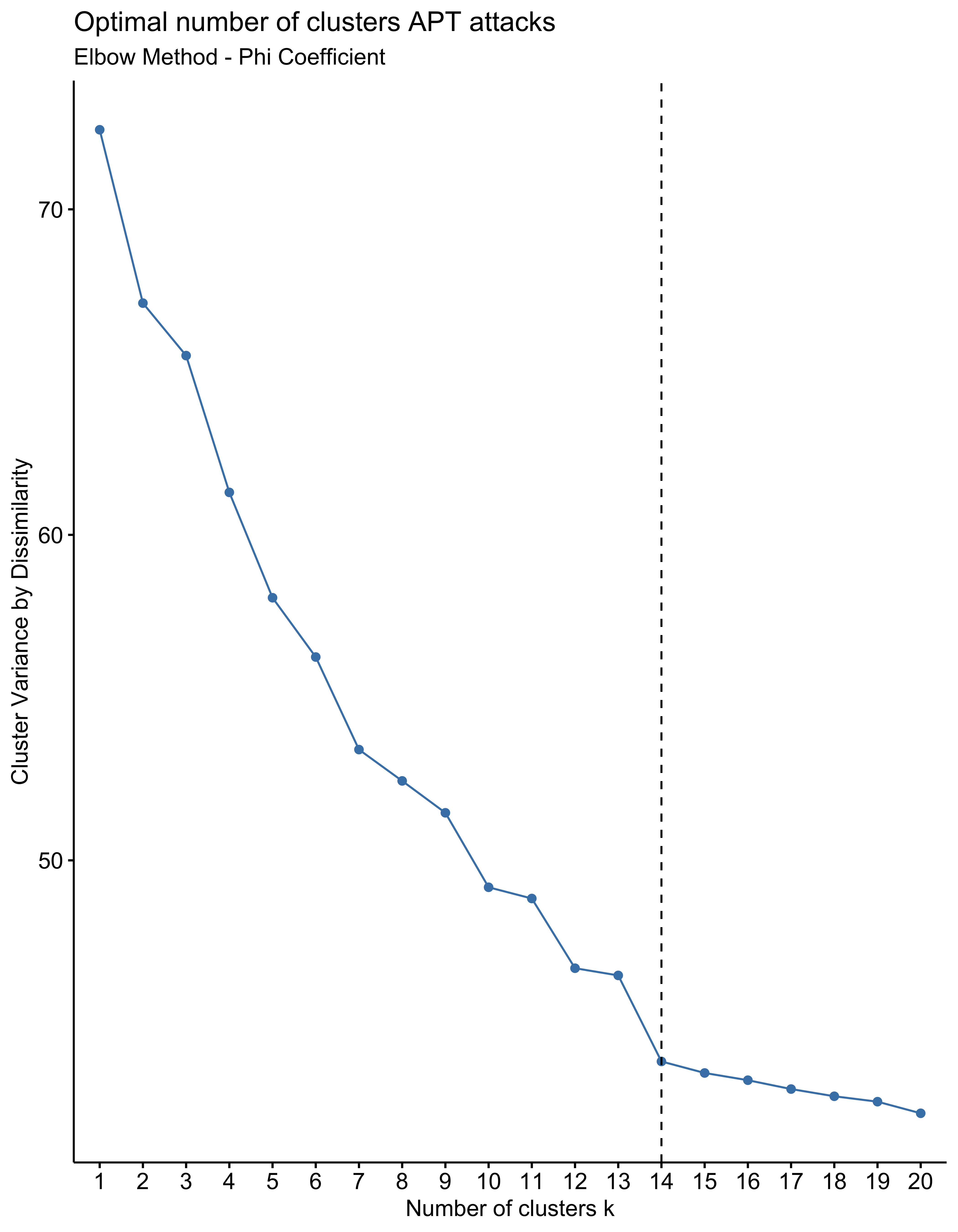}
         \caption{Elbow method}
         \label{fig:KElbowAPT}
     \end{subfigure}
     \hfill
     \begin{subfigure}[b]{0.23\textwidth}
         \centering
         \includegraphics[width=\textwidth, height=3cm, trim={0 0 0 12cm}, clip]{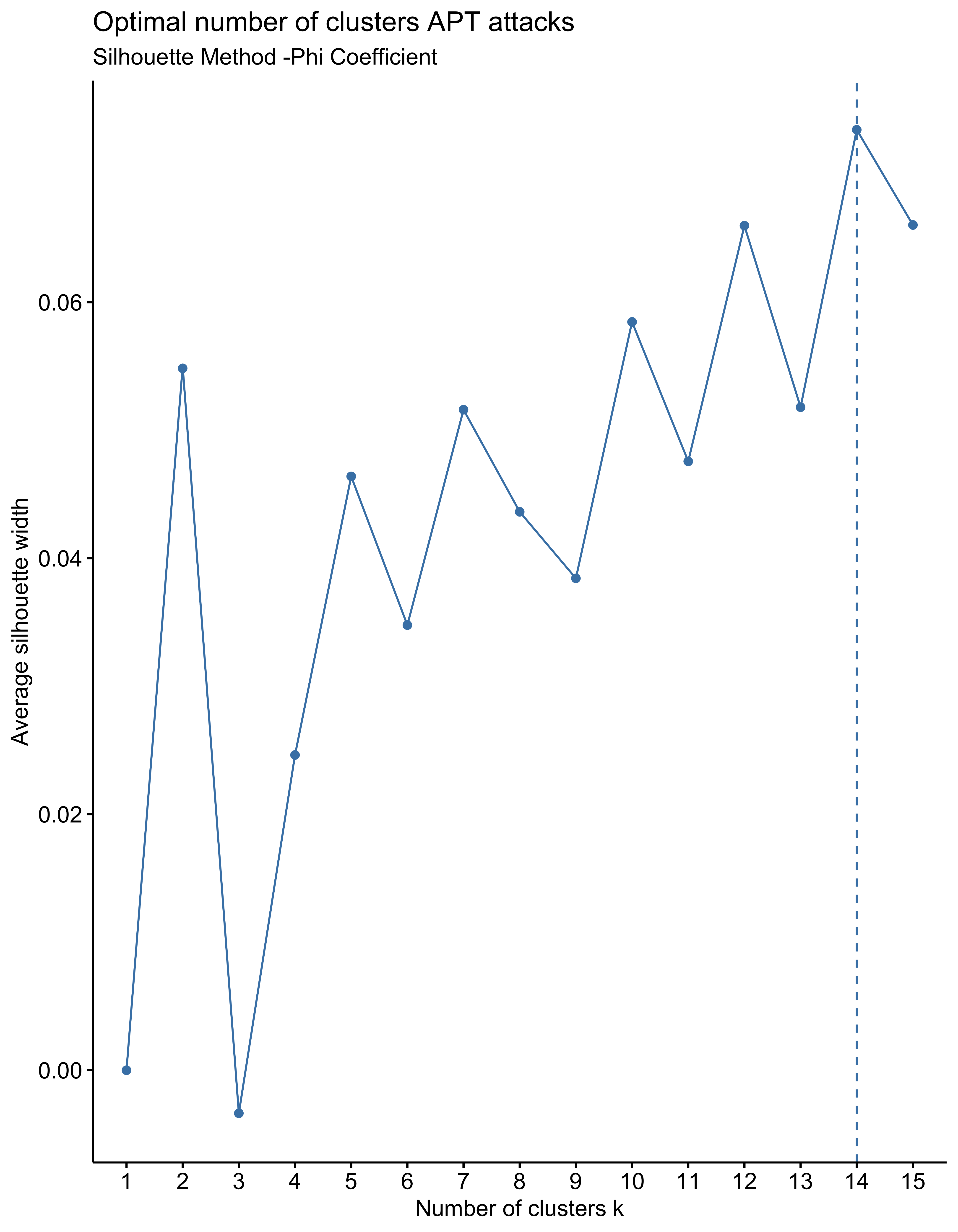}
         \caption{Silhouette method}
         \label{fig:KSilhouetteAPT}
     \end{subfigure}
     \caption{$K$ for APT attacks}
     \label{fig:KAPT}
\end{figure}

\vspace{-0.15em}

\begin{figure}[t]
     \centering
     \begin{subfigure}[b]{0.23\textwidth}
         \centering
         \includegraphics[width=\textwidth, height=3cm, trim={0 0 0 12cm}, clip]{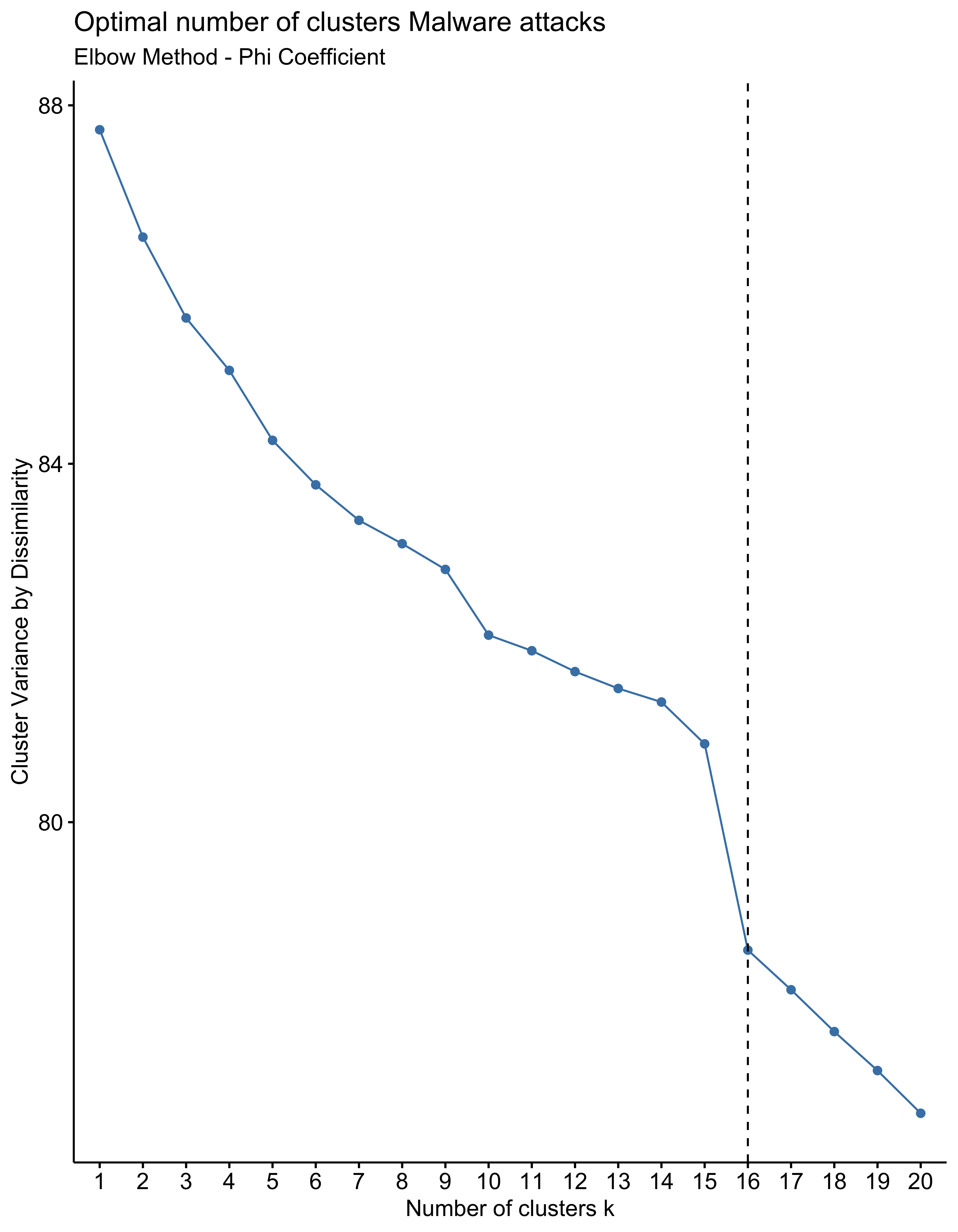}
         \caption{Elbow method}
         \label{fig:KElbowSoftware}
     \end{subfigure}
     \hfill
     \begin{subfigure}[b]{0.23\textwidth}
         \centering
         \includegraphics[width=\textwidth, height=3cm, trim={0 0 0 12cm}, clip]{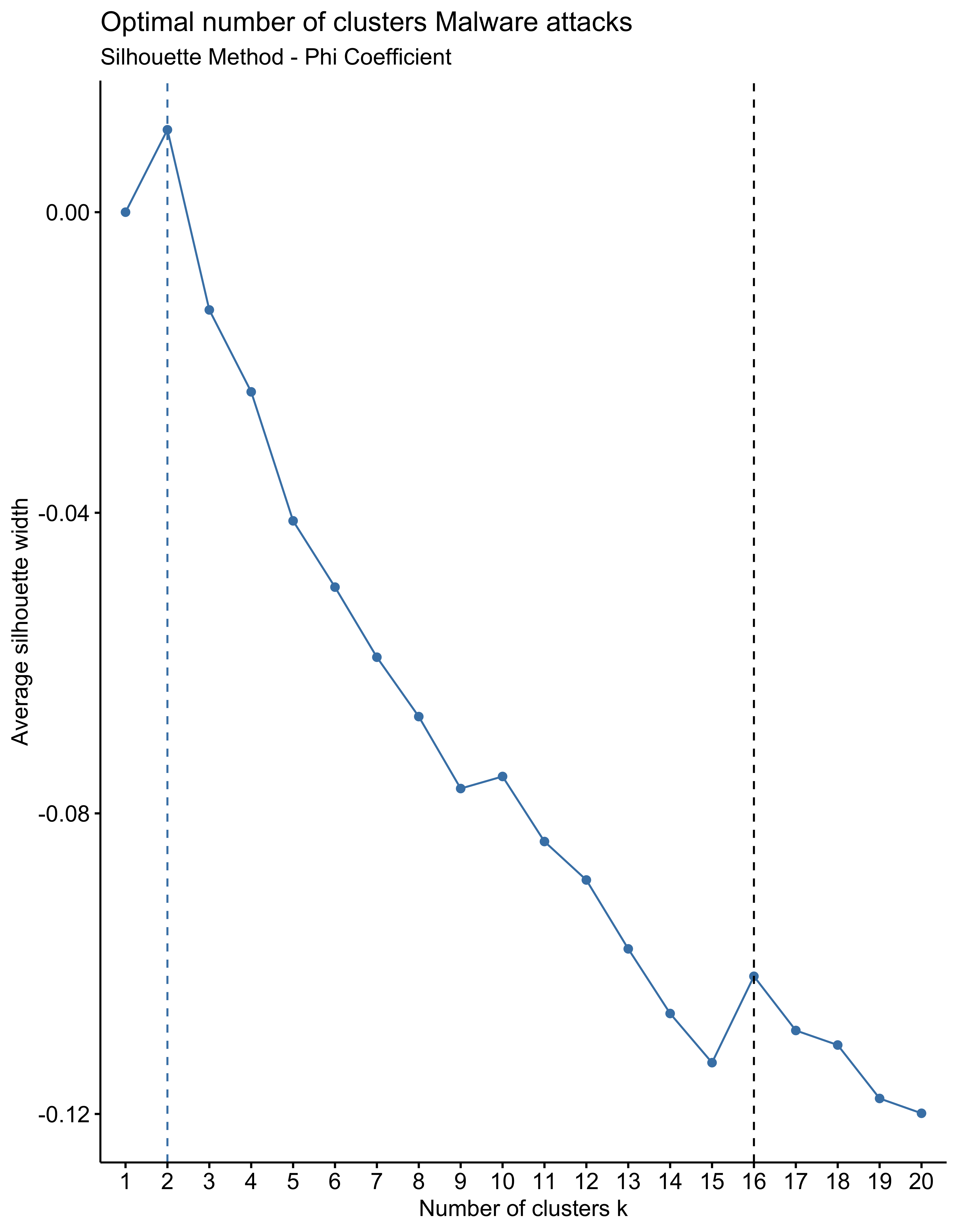}
         \caption{Silhouette method}
         \label{fig:KSilhouetteSoftware}
     \end{subfigure}
     \caption{$K$ for Software attacks}
     \label{fig:KSoftware}
\end{figure}
The Elbow and Silhouette methods are portrayed for both datasets in Figures \ref{fig:KAPT} and \ref{fig:KSoftware}. Observe that both methods resulted in $K=14$ for the dataset of APT attacks. However, for the dataset of Software attacks, the Elbow method determined $K=16$, while the Silhouette method found $K=2$ as the first optimal and $K=16$ as the second optimal. Therefore, we decided to use $K=16$ as the optimal number of clusters for the Software attacks dataset.

\subsubsection{Clustering Algorithms}
\label{Partitioned Algorithms}
The three partitioned clustering methods considered in the preliminary analysis are K-means, PAM, and Fuzzy Clustering.
 
K-means is one of the most common algorithms for clustering a dataset into $K$ clusters. The objective is to minimize the total-within cluster variation, defined as the sum of squared Euclidean distances between items and their corresponding centroid~\cite{ClusteringBook}. The total within-cluster sum of squares for a specific cluster $C_{k}$ will be denoted as $W_{ss}^k$, and is formally expressed as
\begin{equation}
    {W_{ss}^k = {\sum _{x_i\in C_k}{(x_i-\mu_k)^2}}} , \ \ k=1,\dots,K,
\end{equation}
where $x_i$ represents a data point belonging to the cluster $C_k$, and $\mu_k$ is the centroid and mean value of all the points assigned to cluster $C_k$. We deemed K-means clustering as inappropriate for the datasets for the following two reasons: (1) the total within-cluster sum of squares computation uses Euclidean distances, a distance metric inappropriate for the binary datasets; see Remark \ref{Unsuitable Distance Metrics} for details; and (2) the way the centroid is defined in K-means clustering creates a value that cannot be mapped to the original dataset. Moreover, finding the difference between binary points and a mean centroid will indeed result in all the points with a value of 1 to be tied, and all the points with a value of 0 to be tied. K-means clustering does not address ties, and in fact will arbitrarily assign points to a cluster~\cite{KmeansIBM}.

PAM, which is sometimes referred to as K-medoids, is built upon choosing cluster medoids that represent every cluster. Cluster medoids correspond to the most centrally located data point in a cluster. The objective of PAM clustering is to minimize the sum of the dissimilarities between every data point in a cluster and the cluster medoid. The use of medoids as actual data points rather than centroids prompts PAM clustering as a more robust partitioned clustering algorithm to outliers.  Therefore, PAM clustering is less sensitive to noise in the data, compared to K-means clustering~\cite{ClusteringBook}. In addition, the dissimilarity measure for PAM clustering can utilize any distance metric specified, allowing the flexibility to use the appropriate distance metric for the datasets considered in this work.

Finally, Fuzzy clustering is different than K-means and PAM clustering in the aspect that cluster memberships are rather probabilistic. Using a degree of fuzziness, cluster membership probabilities are computed by incorporating that value with the dissimilarity between every point and the medoid. Fuzzy clustering can be visualized by taking the cluster with the highest probability for each data point. Note that, the Fuzzy clustering used in this work follows the Fuzzy analysis clustering algorithm, in contrast with the Fuzzy C-means algorithm which follows K-means centroids and Euclidean distances. Fuzzy analysis clustering allows us to specify the suitable distance metrics for the datasets.

\subsubsection{Clustering Validation}
\label{Clustering Validation}
We validated the performance of partitioned clustering by measuring the intra-cluster compactness and inter-cluster separation using a silhouette analysis and the Dunn index. A silhouette analysis after clustering measures the proximity each technique in one cluster to a technique in its neighboring clustering. A silhouette width near one indicates that the object is well-clustered and separated from neighboring clusters. A silhouette width less than 0 suggests that the object is not well-clustered and is lying in overlapping clusters or between clusters. The average silhouette width is simply an average of the silhouette width for each object clustered, and measures the overall performance of partitioned clustering on a dataset. The Dunn index computes the ratio of the diameter of the clusters and the distance between clusters, and should be maximized for a well-clustered dataset. 

\subsection{Preliminary Results}
\label{preliminary results}
\begin{figure}[t]
     \centering
     \begin{subfigure}[b]{0.23\textwidth}
         \centering
         \includegraphics[width=\textwidth, height=4.7cm, trim={0 0 0 7cm}, clip]{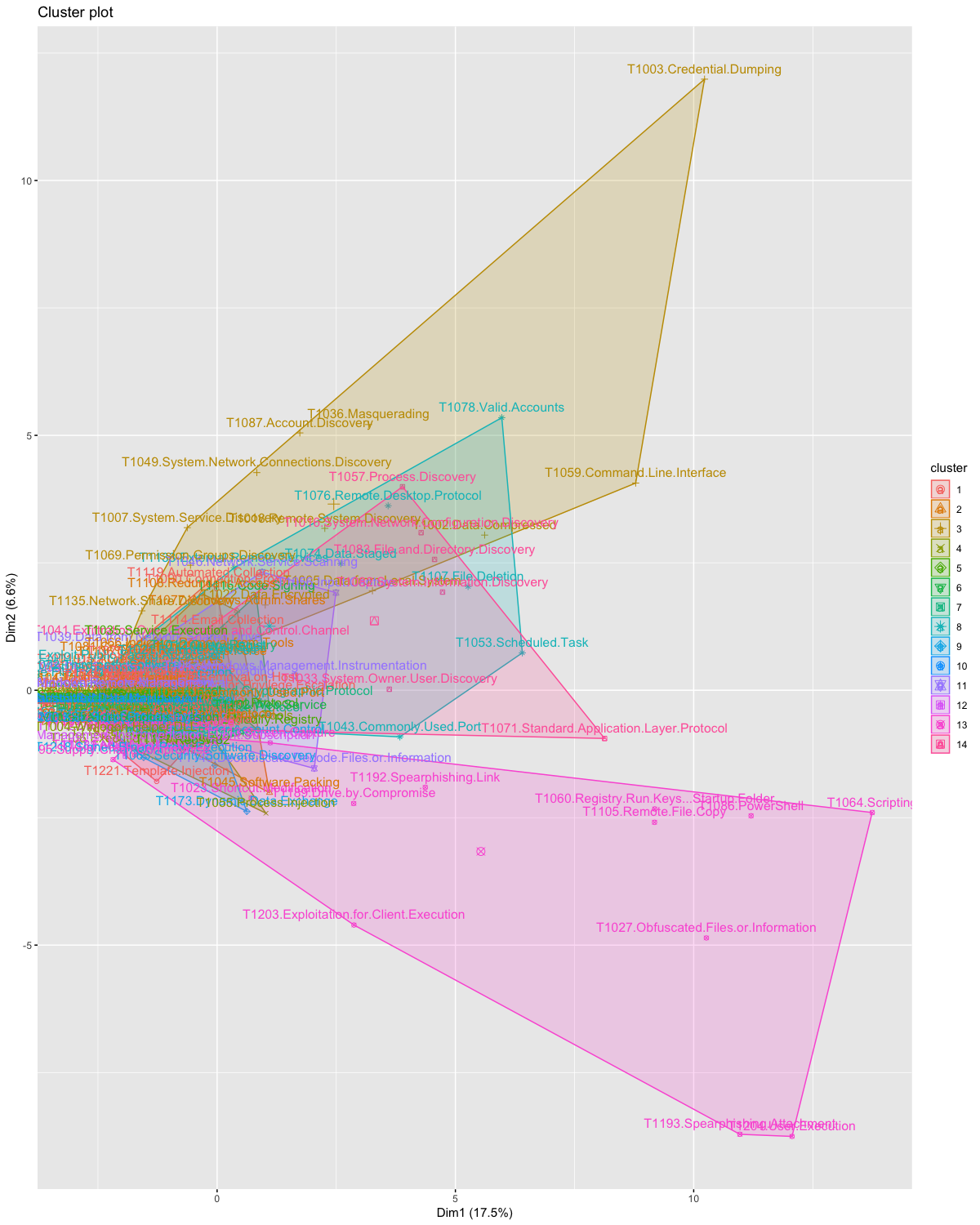}
         \caption{PAM Clustering}
         \label{fig:PAMAPT}
     \end{subfigure}
     \hfill
     \begin{subfigure}[b]{0.23\textwidth}
         \centering
         \includegraphics[width=\textwidth, height=4.7cm, trim={0 0 0 7cm}, clip]{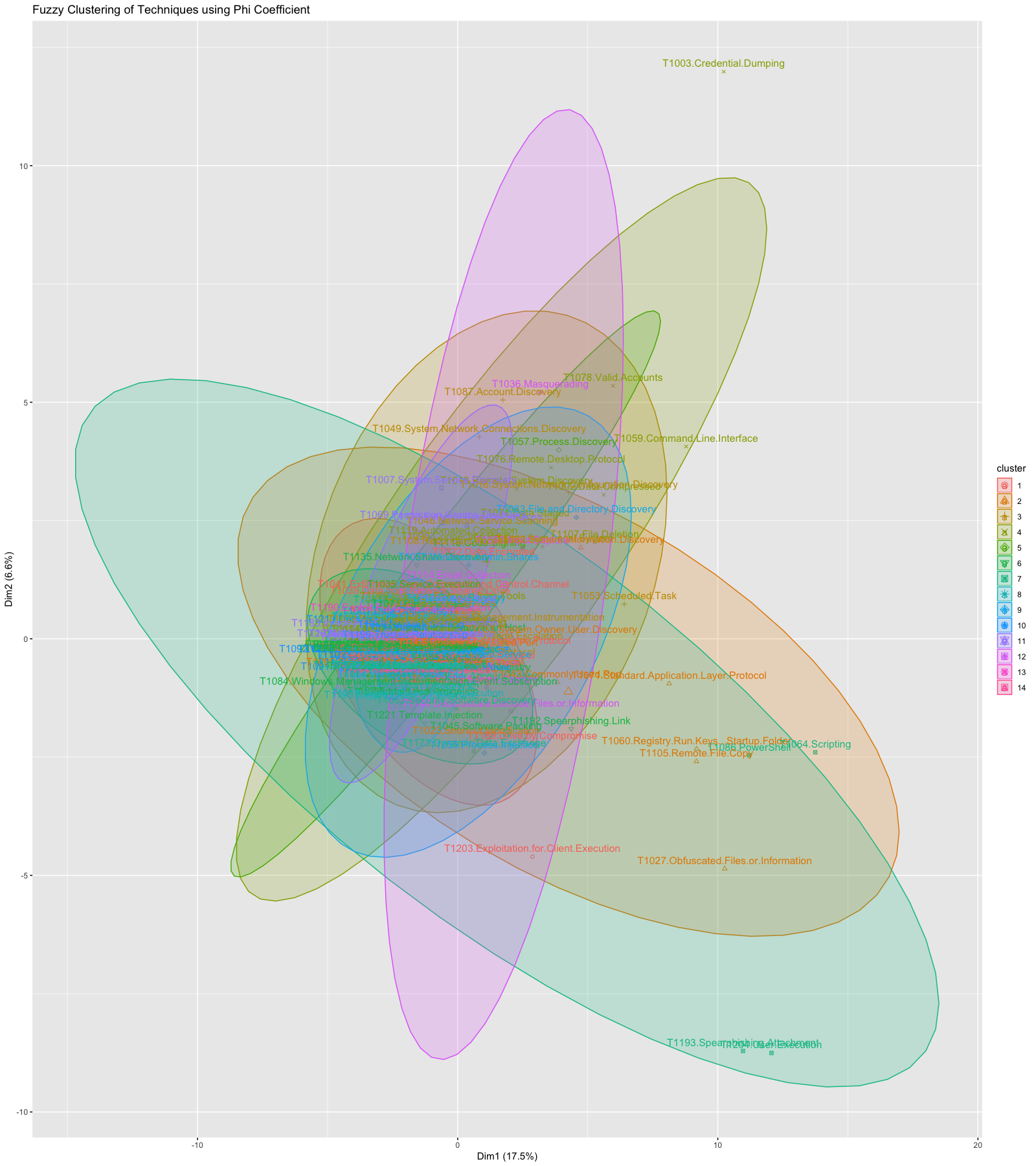}
         \caption{Fuzzy Clustering}
         \label{fig:FuzzyAPT}
     \end{subfigure}
     \caption{Partitioned Clustering APT attacks}
     \label{fig:PAM}
\end{figure}

\begin{figure}[t]
     \centering
     \begin{subfigure}[b]{0.23\textwidth}
         \centering
         \includegraphics[width=\textwidth, height=4.7cm, trim={0 0 0 7cm}, clip]{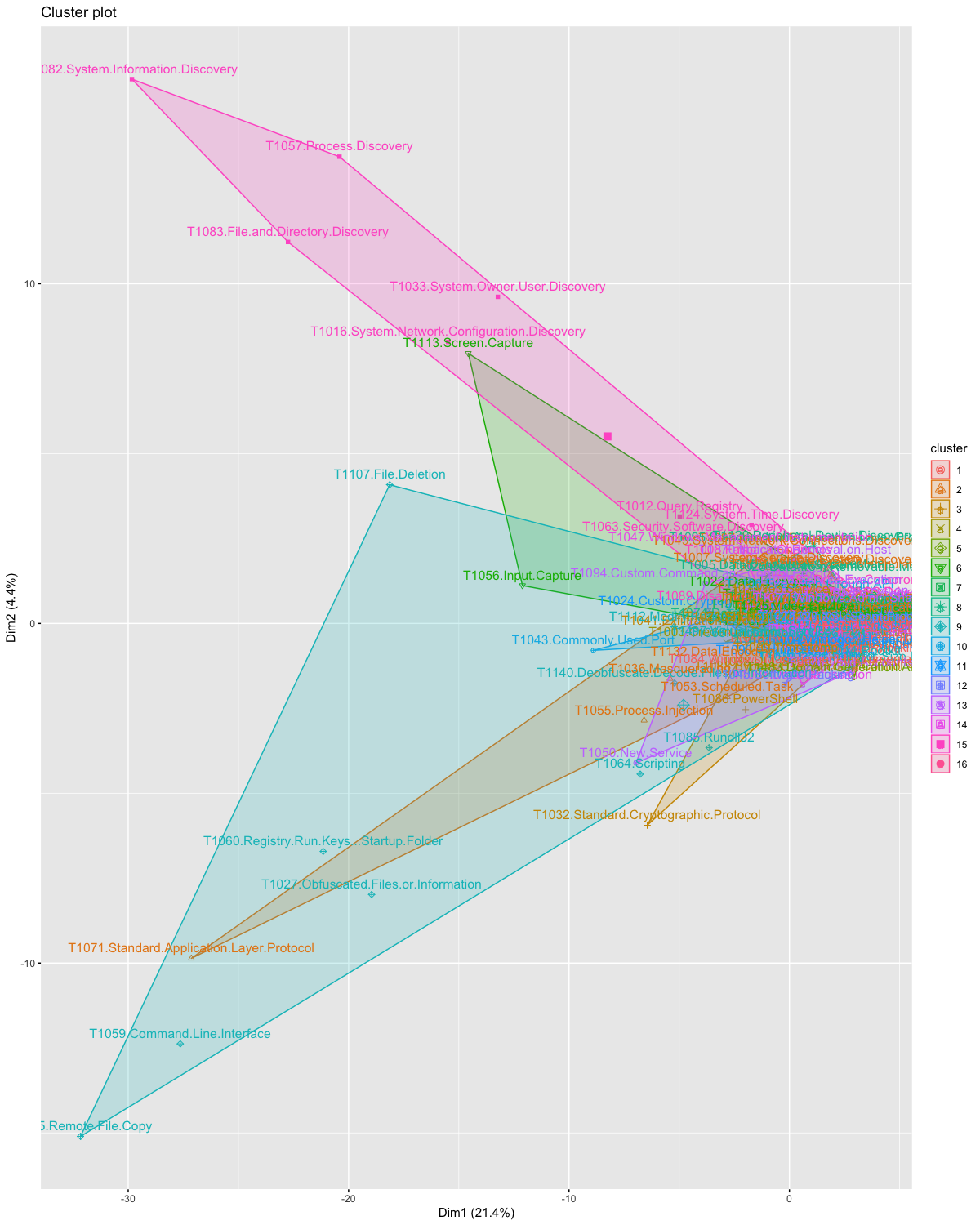}
         \caption{PAM Clustering}
         \label{fig:PAMMalware}
     \end{subfigure}
     \hfill
     \begin{subfigure}[b]{0.23\textwidth}
         \centering
         \includegraphics[width=\textwidth, height=4.7cm, trim={0 0 0 7cm}, clip]{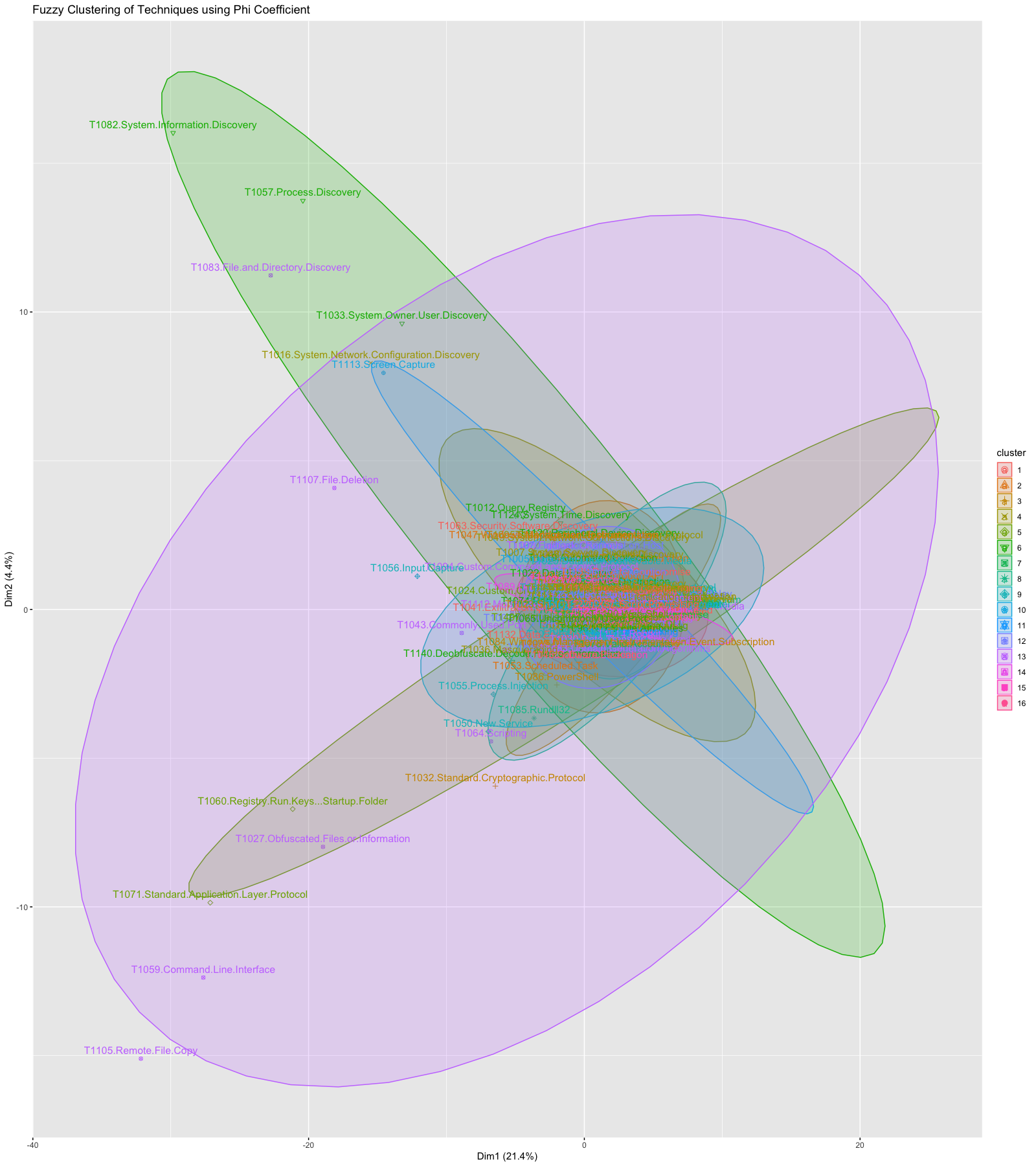}
         \caption{Fuzzy Clustering}
         \label{fig:FuzzyMalware}
     \end{subfigure}
     \caption{Partitioned Clustering Software attacks}
     \label{fig:Fuzzy}
\end{figure}

The cluster plot for each paritioned clustering method and corresponding dataset portrays $K$ clusters represented in different colors, where the axes are reduced to two dimensions for visualization purposes.
The results of PAM and Fuzzy clustering for the APT and Software datasets are shown in Figures \ref{fig:PAM} and \ref{fig:Fuzzy}, respectively. Note that these preliminary clustering results utilized the Phi coefficient correlational distance, discussed in further detail in Section \ref{Distance Metrics}.

\begin{figure}[t]
     \centering
     \begin{subfigure}[b]{0.23\textwidth}
         \centering
         \includegraphics[width=\textwidth, height=4.7cm, trim={0 0 0 5cm}, clip]{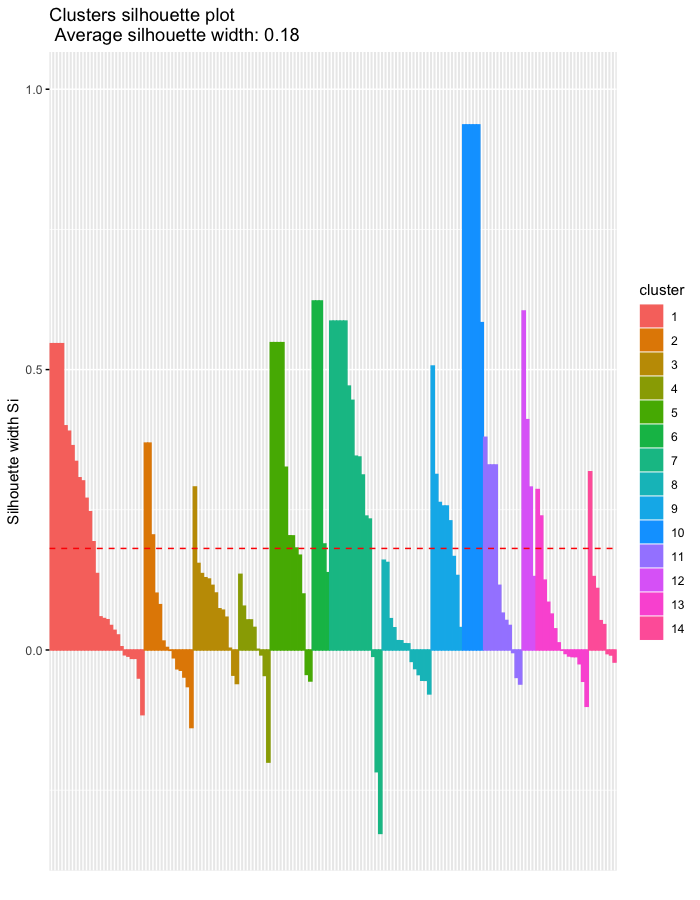}
         \caption{APT attacks }
         \label{fig:SilhouetteAPT}
     \end{subfigure}
     \hfill
     \begin{subfigure}[b]{0.23\textwidth}
         \centering
         \includegraphics[width=\textwidth, height=4.7cm, trim={0 0 0 5cm}, clip]{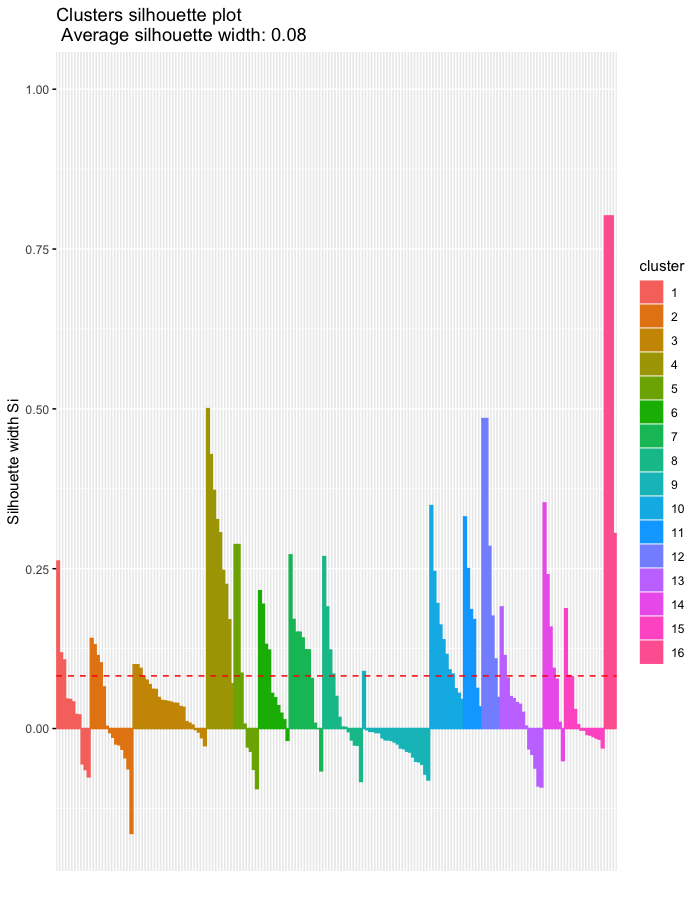}
         \caption{Software attacks}
         \label{fig:SilhouetteSoftware}
     \end{subfigure}
     \caption{Clusters Silhouette Plot for PAM Clustering}
     \label{fig:Silhouette}
\end{figure}

It is evident that the clusters in Figures \ref{fig:PAM} and \ref{fig:Fuzzy} are not well partitioned, and in fact the overlap of the techniques makes it difficult to distinguish any potential technique associations. A silhouette analysis and the Dunn index also confirm that the partitioned clustering of the techniques do not yield accurate or explainable results. 
The average silhouette width for the PAM clustering of the APT and Software attack datasets are $0.18$ and $0.08$, respectively. The clusters silhouette plot is shown in Figure \ref{fig:Silhouette}, where each bar represents a technique in the dataset, and every distinct color is a cluster in the PAM cluster plot. The Dunn index of the PAM clustering for the APT attacks dataset is $0.24$ and for the Software attacks dataset is $0.15$.
Since Fuzzy clustering is based on probabilistic memberships, a silhouette analysis and Dunn index cannot be reported, however, analyzing the technique cluster memberships revealed that the probabilities follow a near uniform distribution. This strongly suggests the lack of well-separated partitions between clusters or interpretability of Fuzzy clustering.

\vspace{-0.25em}

Although the partitioned clustering did not exhibit any clear patterns, we did find some interesting results in the PAM clustering of the APT dataset in Figure \ref{fig:PAMAPT}. For example, techniques \textit{T1193 (Spearphishing Attachment)} and \textit{T1204 (User Execution)} appeared on the bottom right of cluster 13. Interestingly, these two techniques are strongly correlated in a TTP Chain, as specifically described in the MITRE ATT\&CK Framework~\cite{mitre}.
This example indicates that, while there are no clear patterns in the partitioned clustering, there may be a potential for another clustering algorithm that can capture the high dimensional and complex nature of the data. In order to be able to infer accurate and practically meaningful technique associations, our approach, involving hierarchical clustering, is introduced below.

\section{Inferring Attack Technique Associations}
\label{Inferring Attack Technique Associations}
In this section, we will discuss the unique approach we developed in order to infer attack technique associations with a high level of confidence. Section \ref{Distance Metrics} identifies the appropriate distance metric to be used in this work, while Section \ref{Clustering Tendency} discusses the analysis of clustering tendency. Section \ref{Hierarchical Clustering} introduces hierarchical clustering, and Section \ref{Hypothesis Test} presents the statistical hypothesis test performed for inferring the ultimate statistically significant technique associations at a confidence level of 95\%.

\subsection{Distance Metrics}
\label{Distance Metrics}
Clustering is performed on objects using measures of similarity, or distance metrics. As computing distances differs per the nature of the dataset, it is imperative to use the suitable distance measure for the variables in the datasets considered in this work. With the datasets being composed of binary variables, we determined that Jaccard~\cite{jaccard} distances and Phi Coefficient~\cite{phi} correlational distances are the most appropriate.

The Jaccard similarity between any two techniques $T_i$ and $T_j$ is defined as follows

\begin{equation}
    {J_s(T_i, T_j) = \frac {n_{11}}{n_{01} + n_{10} + n_{11}}} , 
\end{equation}

where $n$ is the total number of attack instances in the datasets and $n_{01}$, $n_{10}$, and $n_{11}$ represent the frequency of attacks corresponding to values of $T_i$ and $T_j$, described in Table \ref{phitable}.
Note that, the value of $n_{00}$ is not considered in the computation of the Jaccard similarity index in equation (2), rather only the co-occurrences are taken into account. The Jaccardian distance between techniques $T_i$ and $T_j$ measures the dissimilarity, and is defined as the complement of the Jaccard similarity, that is, 1-$J_s(T_i, T_j)$. The Jaccardian distance between techniques $T_i$ and $T_j$ can be interpreted as the ratio of their intersection divided by their union. While this metric is appropriate for the nature of the datasets considered here, it may not yield explainable technique associations.

The Phi Coefficient is an empirical non-parametric correlation measure specifically for binary data. The Phi Coefficient correlation between any two techniques $T_i$ and $T_j$ is defined as
\begin{equation}
    {r_\phi(T_i, T_j) = \frac {n_{11}n_{00} - n_{10}n_{01}}{n_{1\cdot } n_{0\cdot } n_{\cdot1 } n_{\cdot0}}},
\end{equation}
where all values of $n$ are defined in Table \ref{phitable}.

    \begin{table}[h!]
    \caption{\label{phitable}Distance Metrics Empirical Values}
    \begin{align*}
        \begin{tabular} {|c|c|c|c|}
         \hline
          & \textbf{$\mathbf{T_j}$ = 1} & \textbf{$\mathbf{T_j}$ = 0} & \textbf{total} \\
         \hline
         \textbf{$\mathbf{T_i}$ = 1} & $n_{11}$ & $n_{10}$ & $n_{1\cdot}$ \\
         \hline
         \textbf{$\mathbf{T_i}$ = 0} & $n_{01}$ & $n_{00}$ & $n_{0\cdot}$ \\
         \hline
         \textbf{total} & $n_{\cdot1}$ & $n_{\cdot0}$ & $n$ \\
        \hline
        \end{tabular}
    \end{align*}
    \end{table}
As expressed in equation (3), $r_\phi(T_i, T_j)$ measures the frequency of the co-occurrences and co-absences of techniques $T_i$ and $T_j$. The Phi Coefficient distance between techniques $T_i$ and $T_j$ is defined as the complement of the correlation, that is 1-$r_\phi(T_i, T_j)$. Since the Phi Coefficient is a correlational metric between $T_i$ and $T_j$, it provides the capability of deducing whether techniques $T_i$ and $T_j$ are correlated at a certain level due to their co-occurrences and co-absences in the TTP Chains. Therefore, the Phi Coefficient metric is suitable for the binary datasets and provides an easily interpretable conclusion.

\newtheorem{remark}{Remark}
\begin{remark}[Unsuitable Distance Metric]
\label{Unsuitable Distance Metrics}
One of the most commonly used distance metrics applied for clustering is the Euclidean distance.
Although the Euclidean distance metric is typically applied for continuous variables, it is not appropriate for a binary dataset.
The Euclidean measure drives the calculated distance for a binary dataset to most often be tied and, thus, causes the loss of distinct measures of similarity of the variables. 
\end{remark}

\subsection{Assessing Clustering Tendency}
\label{Clustering Tendency}
Before performing clustering on the datasets, we measured the clusterability of the data. This process, known as assessing clustering tendency, involves evaluating whether the dataset contains meaningful clusters.
This is important because clustering methods can often return clusters even in the absence of notable groups in the dataset. In other words, unknowingly applying a clustering method to a dataset will divide the data into clusters, regardless if they are relevant~\cite{ClusteringBook}. In addition, assessing clustering tendency using the two appropriate distance metrics, Jaccard and Phi Coefficient, will determine which metric is the most suitable for further cluster analysis.

The approach used in this work, known as the Hopkins statistic~\cite{hopkins}, is defined as the probability a given dataset $D$ is generated by a random data distribution. The idea is to compare the distance between the points in dataset $D$ to the distance between points drawn from a randomly simulated dataset $D_R$. Here, $D$ is defined to be either the dataset of APT attacks or Software attacks.
We calculated the Hopkins statistic for the datasets as described in Algorithm \ref{alg:Hopkins}.
Let $\mathbf{H_J}$ denote the Hopkins statistic computed using Jaccard distances, and $\mathbf{H_\phi}$ denote the Hopkins statistic computed using Phi Coefficient correlational distances.
Note that, Step 6 in Algorithm \ref{alg:Hopkins} involves generating $D_R$ using a Bernoulli distribution. This agrees with the nature of the binary variables considered here.

\begin{algorithm}[t]
\caption{Hopkins Statistic Algorithm}\label{alg:Hopkins}
\begin{algorithmic}[1]
\Procedure{Hopkins}{$D$, $m$}
\State Sample uniformly $m$ points $(T_{1}, ...., T_{m})$ from given dataset $D$.
\ForAll {$T_{i} \in D$}
\State $d_i\gets dist(T_i, T_j)$, where $T_j$ is the nearest neighbor \Comment{Compute the distance between $T_i$ and $T_j$ using Jaccard or Phi Coefficient distances}
\EndFor
\State Generate a simulated dataset $D_R$ from a random Bernoulli distribution with $m$ points $(R_1, ...., R_m)$ with the same variance as the given dataset $D$.
\ForAll {$R_{i} \in D_R$}
\State $\tilde{d_i}\gets dist(R_i, R_j)$, where $R_j$ is the nearest neighbor \Comment{Compute the distance between $R_i$ and $R_j$ using Jaccard or Phi Coefficient distances}
\EndFor
\State Calculate the Hopkins statistic as 
\begin{equation}
\label{hopkins}
    {\textbf{H} = \frac {\sum _{{i=1}}^{{m}}\tilde{d_{{i}}}}{{\sum _{{i=1}}^{{m}}d_{{i}}} + {\sum _{{i=1}}^{{m}}\tilde{d_{{i}}}}}} .
\end{equation}
\EndProcedure
\end{algorithmic}
\end{algorithm}

To further understand $\mathbf{H}$, assume $D$ was uniformly distributed and lacking cluster tendencies. Then, the values of ${\sum _{{i=1}}^{{m}}\tilde{d_i}}$ and ${\sum _{{i=1}}^{{m}}d_{{i}}}$ in equation \ref{hopkins} would expectedly be close to one another, making \textbf{H} = 0.5. Therefore, if \textbf{H} $\approx$ 0.5, we conclude that the dataset $D$ is uniformly distributed and does not contain any meaningful clusters.
However, if clusters are present in $D$, then we anticipate \textbf{H} $\approx 1$, and conclude that performing a cluster analysis would yield meaningful results. 
On the other hand, if clusters are not present in $D$, then \textbf{H} $\approx 0$, and we conclude that the data points in $D$ are considered to be regularly spaced, neither random nor clustered.

We computed the Hopkins statistic for both datasets. For the first dataset of APT attacks, $\mathbf{H_J}$ = 0.51 and $\mathbf{H_\phi}$ = 0.60. For the second dataset of Software attacks, we report $\mathbf{H_J}$ = 0.55 and $\mathbf{H_\phi}$ = 0.63. For both of the datasets, we found the reported value of $\mathbf{H_\phi}$ to be higher than $\mathbf{H_J}$, indicating that using the Phi Coefficient correlational distance results in the datasets having a better clustering tendency. For this reason, all our reported results will be expressed in Phi Coefficient correlational distances.

\subsection{Hierarchical Clustering}
\label{Hierarchical Clustering}
Hierarchical clustering is a different method of clustering in which a specified distance matrix is used as the criteria to create a tree-based representation of the data. The tree-like structure resulting from hierarchical clustering is known as a dendrogram, a multilevel hierarchy of the objects being clustered. The number of clusters $K$ need not be specified in advance, rather cutting the tree at a specified height after clustering allows for generating clusters depending on any desired constraints.

Although there are many hierarchical clustering algorithms, we developed a hierarchical clustering method specifically tailored for the datasets considered here. Our approach extends agglomerative hierarchical clustering with Ward's linkage. Agglomerative hierarchical clustering refers to a bottom-up method, where every object in the dataset begins as a separate cluster, or leaf, in the tree. The leaves are then combined into bigger clusters based on the values of the distance matrix. Agglomerative clustering is best for finding clusters of objects with the greatest similarity as it focuses on complete local information from the dataset during fusion decisions, and, therefore, creating smaller clusters where the objects are the most similar~\cite{IRbook}. During the process of agglomerative clustering, combining multiple larger groups of leaves requires a linkage method. Our linkage method for inferring technique associations uses Ward's linkage method. Ward's linkage calculates the distance between larger clusters by computing the sum of squares of the distances, divided by the product of the cardinality of the two clusters. In addition to Ward's linkage creating more compact clusters during hierarchical clustering, Ward's is less susceptible to noise and outliers within the clustering data~\cite{wards}. 

Other types of hierarchical clustering and linkage methods, such as divisive hierarchical clustering and single and complete linkage, do not possess the same qualities of compactness and robustness as our approach of agglomerative hierarchical clustering with Ward's linkage. Divisive hierarchical clustering refers to a top-down approach at which the objects in the dataset are recursively split into clusters until each object becomes a singleton. During each iteration of division, the split decision is made by comparing the dissimilarity of the objects to one another~\cite{IRbook}. Divisive clustering does not conform with the objective of inferring technique associations that represent correlations between techniques that are at the highest degrees of similarity in attack instances. Similarly, we deemed other well-known linkage techniques such as single linkage and complete linkage as not suitable for our goal. Single-linkage, otherwise known as minimum linkage, creates long-loose clusters, and often cannot separate clusters in the hierarchy in the presence of noise. Complete linkage possesses the tendency to unseeingly break larger clusters, which may does not contribute to our goal of maintaining the relationships between associations~\cite{HclustSlides}. For those reasons, our inference of the technique associations used agglomerative hierarchical clustering with Ward's linkage. The results of divisive hierarchical clustering for the Software and APT attack datasets are reported in Figures \ref{fig:DivAPT} and \ref{fig:DivSoftware} in the Appendix. In addition, a representation of the hierarchical clustering using different linkage methods is portrayed in Figure \ref{fig:ensemble}.

We extended the agglomerative hierarchical Ward's linkage algorithm to incorporate the Phi coefficient correlational distance metric for the APT and Software attacks datasets. Performing this hierarchical clustering will grant the ability of inferring meaningful and accurate technique associations by encompassing clusters that have the highest degrees of correlation.

Although we now have the final dendrogram of the learned hierarchical clustering tree, we are still not able to infer the significant technique associations because we need to determine a cutoff in the tree. Our unique approach to assess the validity and derive the statistically significant cutoff for the technique associations is discussed in the next section.

\subsection{Statistical Hypothesis Testing of Hierarchical Clustering}
\label{Hypothesis Test}
After developing the hierarchical clustering tree, the final steps are, first, to assess the validity of the tree and, second, create a cutoff at a height of the dendrogram in order to create the final clusters. For the datasets, the final clusters learned from the hierarchical clustering tree will represent the technique associations. The novel approach used in this work addresses these two aspects by performing a statistical hypothesis test on the learned agglomerative Ward's linkage hierarchical clustering tree, which will be denoted as $T_D$. Specifically, a statistical hypothesis test will analyze the validity of the clusters in $T_D$ by comparing the learned tree to a tree resulted from a null distribution, and will allow us to infer statistically significant results at the desired confidence level~\cite{StatTestBook}.

The null tree used for the statistical hypothesis test is generated from a random Bernoulli distribution with the same variance as that of the dataset, and will be denoted as $T^0$. If the clusters in $T_D$ are considerably different than that of $T^0$ at the specified cutoff, then the learned tree will yield statistically significant results. An outcome of a statistically significant hierarchical tree grants the conclusion that the resulting associations are far from random chance, providing validity to the technique associations. The approach is further explained in Algorithm \ref{alg:StatTest}.

\begin{algorithm}[t]
\caption{Statistical Test of Hierarchical Clustering}\label{alg:StatTest}
\begin{algorithmic}[1]
\Procedure{Test}{$T_D$}
\For{possible $k\gets 1$ to 100 } 
\State $h_k\gets$ cutoff height of $T_D$ that creates $k$ clusters
\For{$j\gets1$ to 1,000 } 
\State Generate a null tree, $T^0_j$, using agglomerative Ward's linkage from a random Bernoulli distribution with the same variance as that of $T_D$
\State $h^0_{k,j}\gets$ cutoff height for $T^0_j$ that creates $k$ clusters
\EndFor
\State $p\gets$ count $h^0_{k,j} \leq h_k$ / 1,000
\EndFor
\State Find the first value stored in $p$ that is $\leq$  0.05 to conclude statistical significance at a 95\% confidence level.
\EndProcedure
\end{algorithmic}
\end{algorithm}

\subsection{Experimentation and results}
\label{experimentation}
\begin{figure}[t]
     \centering
     \begin{subfigure}[b]{.23\textwidth}
         \centering
         \includegraphics[width=\textwidth, height=4cm, trim={0 0 0 12cm}, clip]{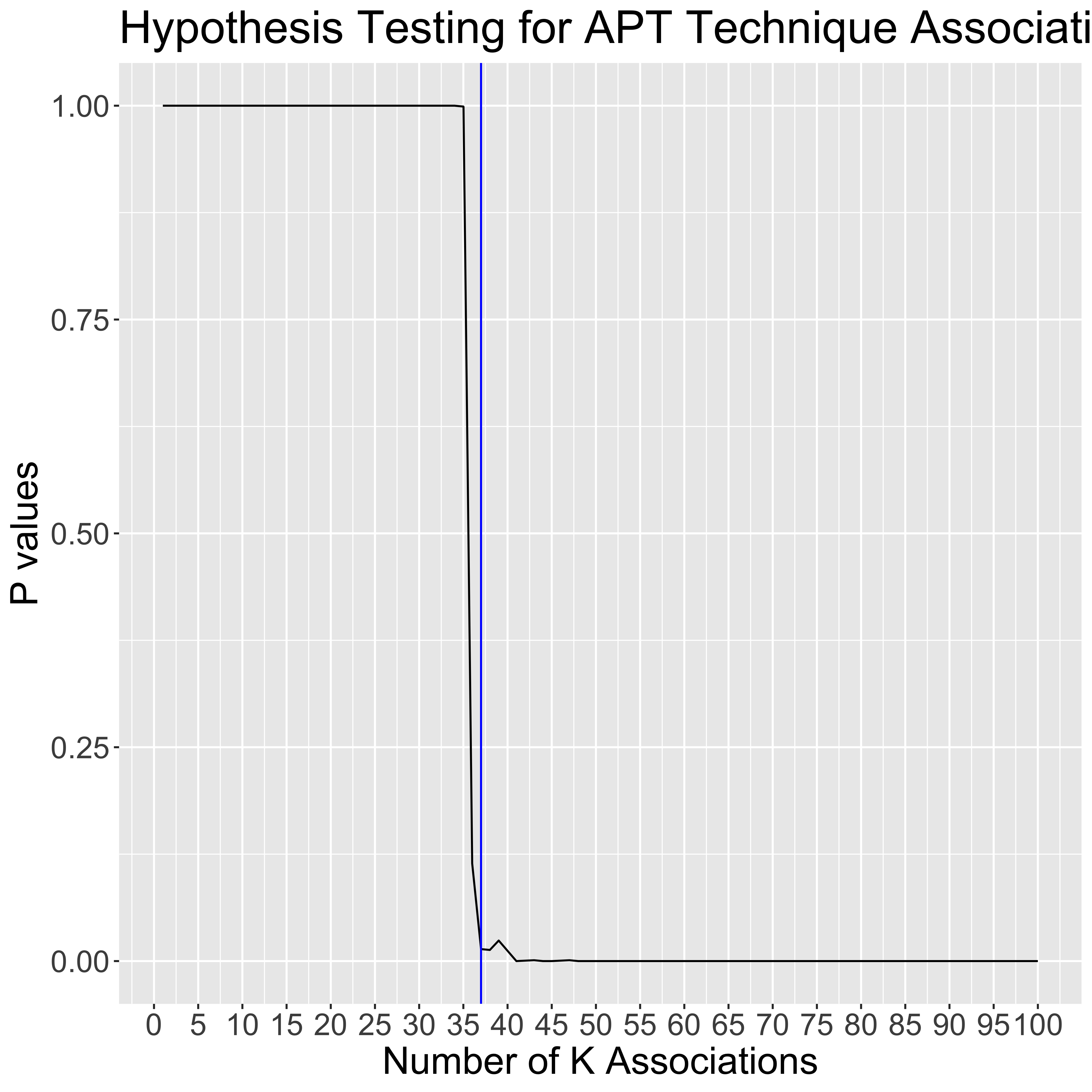}
         \caption{APT attacks}
         \label{fig:TestAPT}
     \end{subfigure}
     \hfill
     \begin{subfigure}[b]{.23\textwidth}
         \centering
         \includegraphics[width=\textwidth, height=4cm, trim={0 0 0 12cm}, clip]{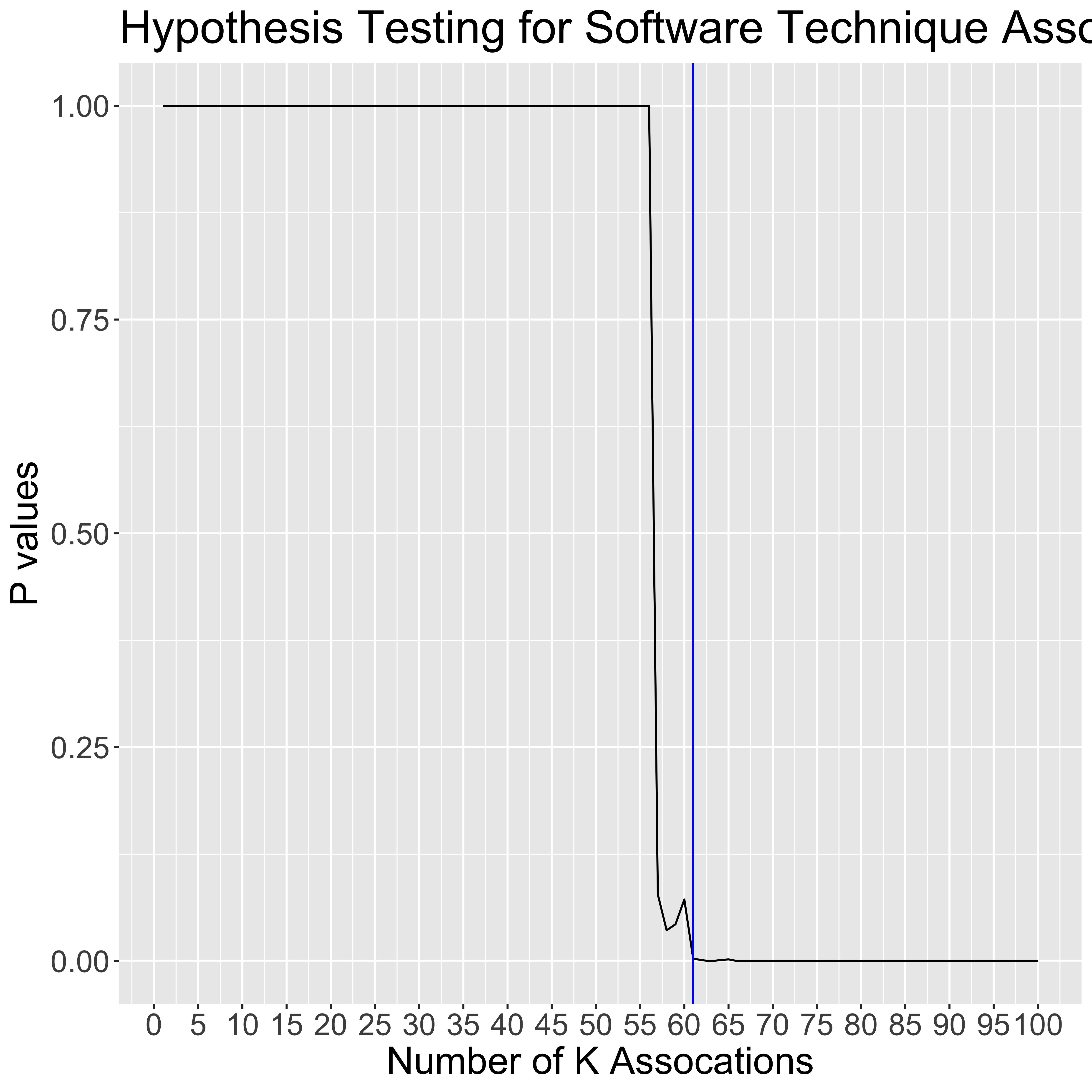}
         \caption{Software attacks}
         \label{fig:TestSoftware}
     \end{subfigure}
     \caption{Hypothesis Test of Learned Hierarchical Clustering}
     \label{fig:Hypothesis Test}
\end{figure}

First, we determined the cutoff value of the agglomerative Ward's clustering tree for the dataset of APT attacks and Software attacks at a rigorous 95\% confidence level. The resulted $p$-values are shown in Figure \ref{fig:Hypothesis Test}. The blue vertical line in Figures \ref{fig:TestAPT} and \ref{fig:TestSoftware} correspond to the statistically significant amount of associations, 37 and 61 associations for the APT and Software attacks, respectively.

\begin{figure*}[h!]
  \includegraphics[scale=0.12, trim={0 0 0 3cm}, clip]{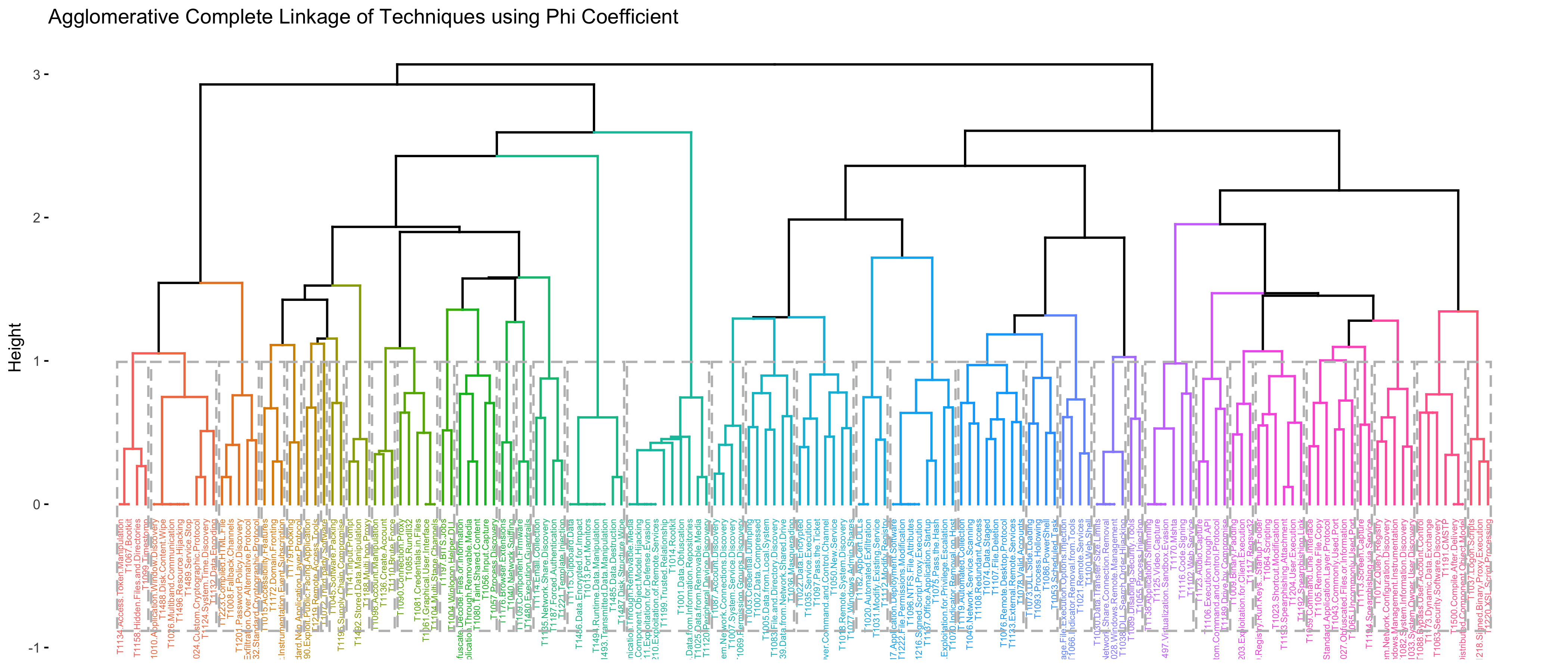}
  \caption{Agglomerative Hierarchical Clustering of APT Attacks}
  \label{fig:HclustAPT}
\end{figure*}

\begin{figure*}[h!]
  \includegraphics[scale=0.12, trim={0 0 0 3cm}, clip]{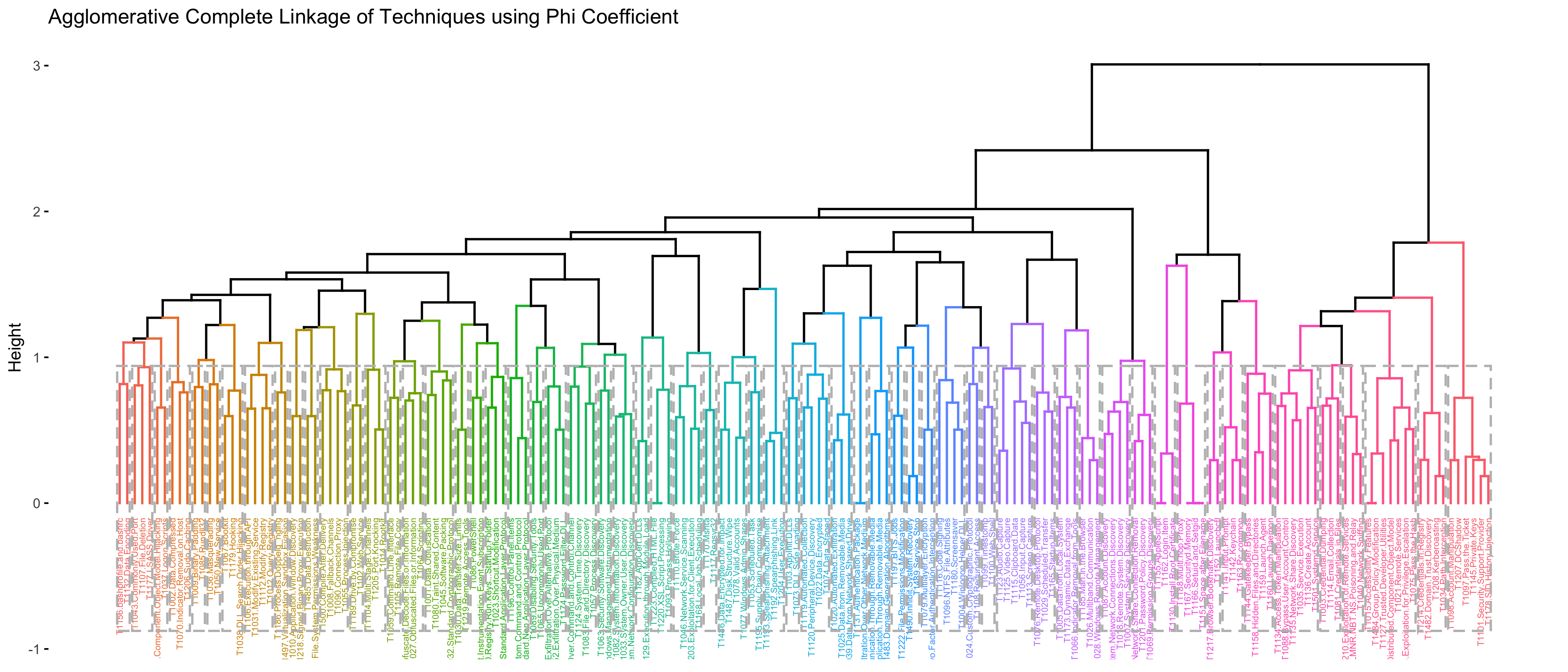}
  \caption{Agglomerative Hierarchical Clustering of Software Attacks}
  \label{fig:HclustSoftware}
\end{figure*}

The results of the learned hierarchical clustering tree for the dataset of APT attacks is portrayed in Figure \ref{fig:HclustAPT} and for the dataset of Software attacks, Figure \ref{fig:HclustSoftware}.
Each cluster, or fine-grain association, in the figure is surrounded by a gray dashed box, and is represented in a different color in the tree. 

\section{Evaluation}
\label{Evaluation}
In addition to the statistical hypothesis test providing validity to the learned hierarchical clustering presented in Section \ref{Hypothesis Test}, in this section we will present two methods to evaluate the accuracy of the learned technique associations. 

\subsection{Measuring Mutual Information}
We evaluated the learned attack technique associations using an information theoretic approach. We computed the mutual information of the techniques in the fine-grain clusters, as well as the coarse-grain clusters directly from the datasets. Mutual information allows us to evaluate the learned correlations by identifying the relatedness or dependence between any two techniques within the same cluster (fine-grain) or across joint clusters (coarse-grain) independent of the assumptions of the underlying probability distributions~\cite{mi}. In order to be able to compare the mutual information between different associations, we performed our evaluation using the normalized mutual information (NMI)~\cite{nmi}.

\begin{figure}[t]
     \centering
     \begin{subfigure}[b]{.23\textwidth}
         \centering
         \includegraphics[width=\textwidth]{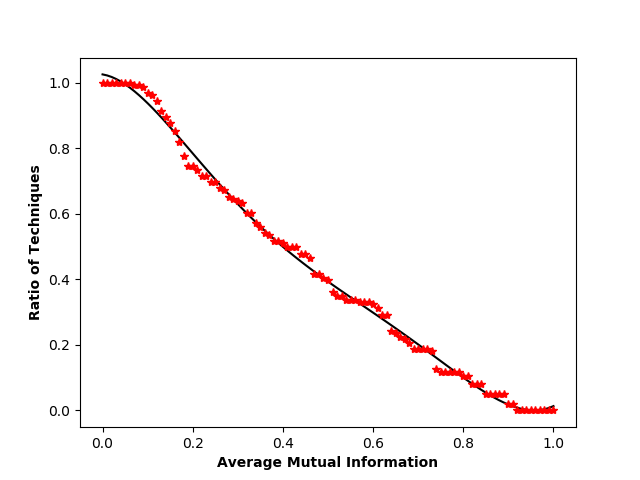}
         \caption{Technique-based NMI Measure}
         \label{fig:techniqueNMIAPT}
     \end{subfigure}
     \hfill
     \begin{subfigure}[b]{.23\textwidth}
         \centering
         \includegraphics[width=\textwidth]{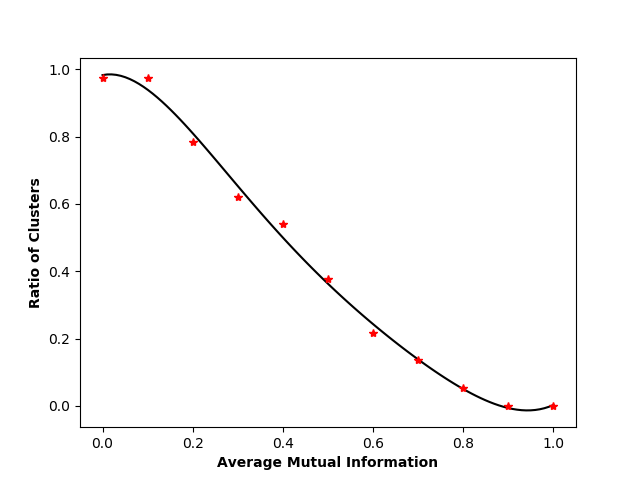}
         \caption{Cluster-based NMI Measure}
         \label{fig:clusterNMIAPT}
     \end{subfigure}
     \caption{Analysis of Fine-grain Associations for APT Attacks}
     \label{fig:FinegrainAPT}
\end{figure}

\begin{figure}[t]
     \centering
     \begin{subfigure}[b]{.23\textwidth}
         \centering
         \includegraphics[width=\textwidth]{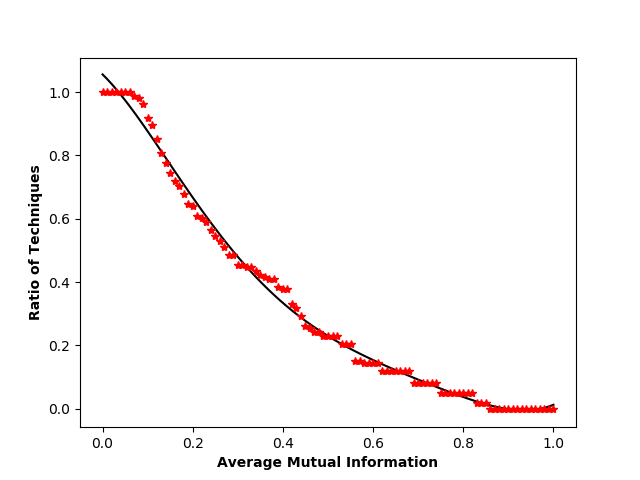}
         \caption{Technique-based NMI Measure}
         \label{fig:techniqueNMISoftware}
     \end{subfigure}
     \hfill
     \begin{subfigure}[b]{.23\textwidth}
         \centering
         \includegraphics[width=\textwidth]{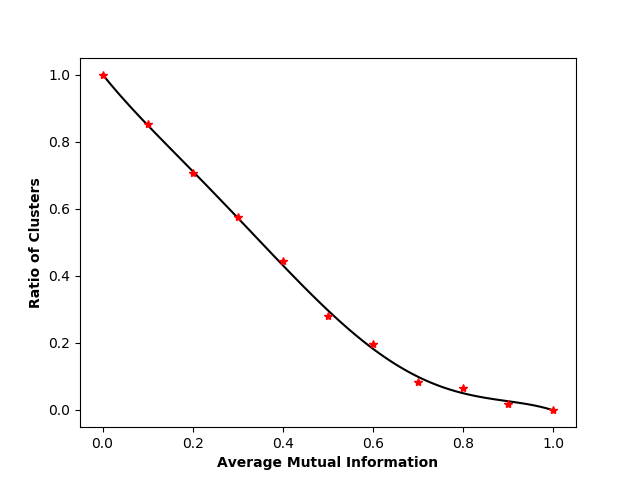}
         \caption{Cluster-based NMI Measure}
         \label{fig:clusterNMISoftware}
     \end{subfigure}
     \caption{Analysis of Fine-grain Associations for Software Attacks}
     \label{fig:finegrainSoftware}
\end{figure}

First, we performed our analysis of the fine-grain associations for both datasets. This involved a technique-based NMI measure as well as a cluster-based NMI measure. We computed the technique-based measure by calculating the NMI for every pairwise combination of techniques in the same cluster. This quantitatively represents the maximum predictability each technique possesses based on its cluster assignment. As for the cluster-based measure, we computed the NMI for each technique in the cluster and found the average, yielding a measure of how predictable that cluster is. We identified the threshold of the NMI value for the fine-grain analysis by empirically assessing the technique occurrences in the datasets. We found that an NMI value of 0.25 is equivalent to having a co-occurrence of 75\%, which shows that the learned correlation is practically useful for prediction. The complement cumulative distribution function of the fine-grain analysis is shown for both datasets in Figures \ref{fig:FinegrainAPT} and \ref{fig:finegrainSoftware}. 
The results of the fine-grain analysis show that from the dataset of APT attacks, 78\% of the techniques (Figure \ref{fig:techniqueNMIAPT}) and 75\% of the clusters (Figure \ref{fig:clusterNMIAPT}) indicate high predictability. Similarly for Software attacks, 60\% of the techniques and clusters (Figure \ref{fig:techniqueNMISoftware} and Figure \ref{fig:clusterNMISoftware}) from the learned hierarchical clustering tree indicate high predictability. 

\begin{figure}[t]
     \centering
     \begin{subfigure}[b]{.23\textwidth}
         \centering
         \includegraphics[width=\textwidth]{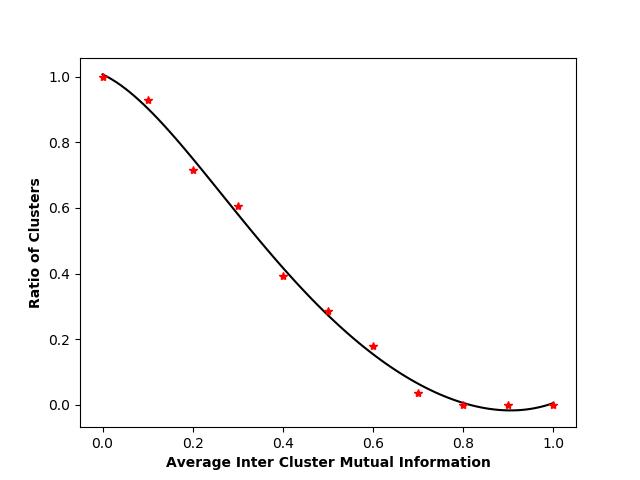}
         \caption{Inter-cluster NMI Measure for APT Attacks}
         \label{fig:interAPT}
     \end{subfigure}
     \hfill
     \begin{subfigure}[b]{.23\textwidth}
         \centering
         \includegraphics[width=\textwidth]{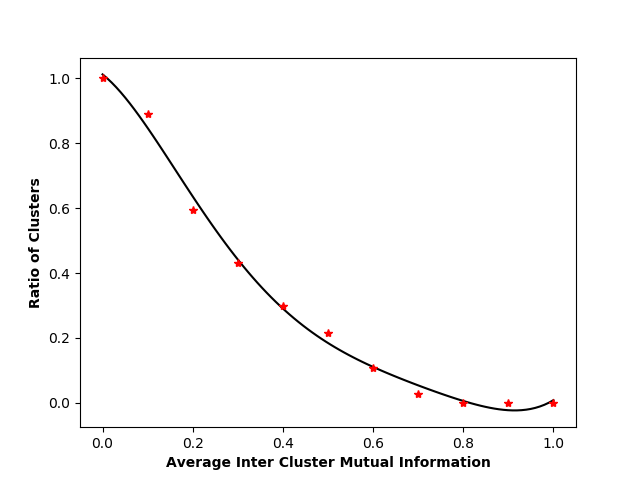}
         \caption{Inter-cluster NMI Measure for Software Attacks}
         \label{fig:interSoftware}
     \end{subfigure}
     \caption{Analysis of Coarse-Grain Association}
     \label{fig:coarsegrain}
\end{figure}

Moreover, we evaluated the coarse-grain associations by computing the inter-cluster NMI for both datasets. We measured the inter-cluster NMI by finding the neighboring connecting cluster in the next level of the hierarchical clustering tree and computing the NMI between the techniques in the connected clusters. The complement cumulative distribution function of the coarse-grain analysis is shown for both datasets in Figure \ref{fig:coarsegrain}. 
The results of the coarse-grain analysis show that, 75\% of the inter-cluster cluster correlations for the APT attacks (Figure \ref{fig:interAPT}) and 60\% of the inter-cluster correlations for the Software attacks (Figure \ref{fig:interSoftware}) indicate high predictability of their connected cluster at the next level of the hierarchy. 

In Section \ref{discussion}, we will discuss examples of fine-grain and coarse-grain associations, as well as the relationships they manifest in further detail.

\subsection{Evaluation Based on Domain Experts}
We also recruited 6 Cybersecurity experts from both academic institutions and government who have at least 5 years of experience in the area of cyber threat intelligence and are familiar with the MITRE ATT\&CK Framework. We provided each expert with an evaluation rubric to label each association as ``agree", ``disagree", or ``neutral" along with the corresponding justifications. The experts confirmed the presence of strong correlations for 93\% of the fine-grain technique associations, and 90\% of the coarse-grain associations. In addition, when investigating the fine-grain and coarse-grain associations that were labeled as ``disagree" or ``neutral", we found those associations difficult to justify since the techniques in those associations have low occurrences in the datasets.

\section{Discussion}
\label{discussion}
In this section we will discuss some findings and association examples based on our clustering analysis. 

\paragraph*{\textbf{Hierarchical clustering reflects the complexities of APT and Software attacks}} The average size of the fine-grain associations of the Software attacks is smaller than that of the APT attacks. This is because the Software attacks typically extend a smaller number of tactics, while APT attacks are typically more complex and span several tactics in the attack chain. 

\paragraph*{\textbf{Fine-grain associations}} There are 37 and 61 fine-grain clusters for APT and Software attacks, respectively. These intra-cluster correlations represent various relationships (conjunctive, disjunctive or sequential) between techniques within the same cluster. Although the clustering itself may not reveal the kind of relationship between techniques, this can be identified by expert inspection based on techniques functions. Will will discuss two examples of this.

\begin{figure}[t]
    \begin{center}
        \includegraphics[width=0.99\linewidth]{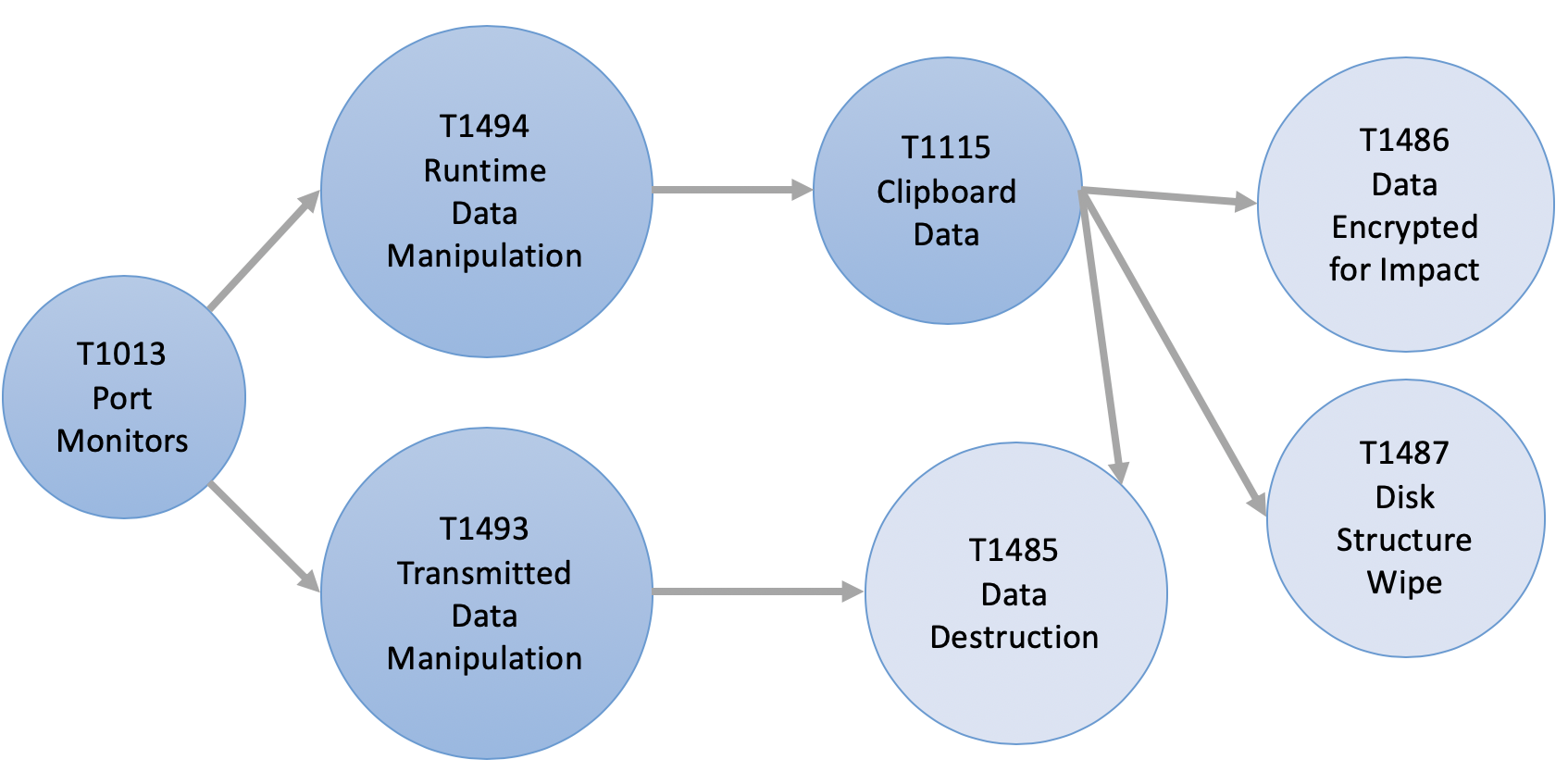}
    \end{center}
  \caption{Fine-grain Association Example}
  \label{fig:example}
\end{figure}

In the first example, the cluster composed of \{\textit{T1013 (Port Monitors)}, \textit{T1494 (Runtime Data Manipulation)}, \textit{T1493 (Transmitted Data Manipulation)}, \textit{T1115 (Clipboard Data)}, \textit{T1485 (Data Destruction)}, \textit{T1486 (Data Encrypted for Impact)}, and \textit{T1487 (Disk Structure Wipe)}\} represents an attack pattern for accomplishing data destruction as an adversarial goal. 
The adversary starts with \textit{T1013} to load malicious code at startup through a DLL. Then, the adversary may choose to perform \textit{T1494} or \textit{T1493} to manipulate runtime or transmitted data to negatively affect a business process. Afterwards, the adversary may collect different sources of data, such as \textit{T1115} that can later impact data availability. Finally, the adversary performs one or more of \textit{1485}, \textit{T1486}, or \textit{T1487} to cause data unavailability. Attackers usually plan for multiple techniques (from the tactic) that are performed concurrently (cognitively) or selectively (as alternatives) in order to maximize the potential of success. 
Figure \ref{fig:example} shows the relationship manifested in this association. %

As a second example, the \{\textit{T1028 (Windows Remote Management)}, \textit{T1038 (DLL Search Order Hijacking)}, \textit{T1030 (Data Transfer Limits)}, and \textit{T1126 (Network Share Connection Removal)}\} association represents an attack pattern in the TTP Chain of an information stealer (e.g., \textit{Threat Group 3390}~\cite{mitre}). First, the adversary executes malicious code leveraging the Windows Remote management protocol as enabled by technique \textit{T1028}. As a result, the adversary can then replace or modify a DLL to gain privilege escalation and persistence using (\textit{T1038}). Afterwards, the adversary will employ technique \textit{T1030} to exfiltrate collected data in fixed size chunks to avoid data transfer alerts. Finally, the adversary cleans up the traces of their operation by detaching the network shares, (\textit{T1126}), after exfiltration for defense evasion. Evidently, the relationship in this association is sequential.

\paragraph*{\textbf{Coarse-grain associations}} The fine-grain associations are combined together at various levels to form bigger clusters in the hierarchy. We call this inter-cluster association coarse-grain. There are 35 first-level coarse-grains associations in the APT attack hierarchical clustering tree. As we observed, the coarse-grain associations show complementary techniques performed across various tactics in the TTP Chain. For instance, the second example association discussed above (\{\textit{T1028 (Windows Remote Management)}, \textit{T1038 (DLL Search Order Hijacking)}, \textit{T1030 (Data Transfer Limits)}, and \textit{T1126 (Network Share Connection Removal)}\}) is connected to another association \{\textit{T1055 (Process Injection)} and \textit{T1089 (Disabling Security Tools)}\} at the next level of the hierarchy. 
An attacker may use the latter cluster  to obtain defense evasion during the former cluster. 
That is, \textit{T1055} for executing code masked under a legitimate process and \textit{T1089} for disabling security tools are both used to avoid detection. For an APT, this complementary relationship is critical to their goal, and the coarse-grain association aids in identifying the larger attack chain.

Interestingly, many of the associations inferred by our approach, as examples discussed above, may not be easily deduced by experts in the field. The fine-grain and coarse-gran associations go beyond what experts can manually discover. Thus, the knowledge from this work can be used to further advance the ability to predict adversarial behavior outside the limits of expertise or heuristics.

\section{Conclusion and Future Work}
Overall, in this work we accurately learn the fine-grain and coarse-grain technique associations based on real-life APT and Software attack datasets. These associations are important because they enable the prediction of adversarial behavior based on observed techniques, which can be directly applied to attack diagnosis and threat mitigation. Our approach first involves establishing the suitable distance metric of the datasets and extending hierarchical clustering algorithms in order to infer explainable fine-grain and coarse-grain associations. Moreover, we deduce statistically significant correlations and assess the validity of our learned hierarchical clustering tree by performing a hypothesis test and ensure that no more than 5\% of the fine-grain associations are due to random chance. Our approach leads to the inference of 37 fine-grain technique associations from the APT attacks dataset and 61 from the Software attacks dataset. Then, we evaluate the learned fine-grain and coarse-grain associations using mutual information, and the results show that 78\% of the fine-grain and 75\% of the coarse-grain technique associations exhibit high predictability for APT attacks. 

Our approach encountered a few limitations that provide avenues for future work.
First, we will work on collecting a larger dataset of real-life attacks in order to increase the number and the quality of technique clusters and associations.
Second, we will investigate an approach, involving fuzzy hierarchical clustering, to explore probabilistic technique associations as well as other methods methods to reveal precondition and post-condition relationships.

\section*{Acknowledgements}
\label{Acknowledgements}
We would like to thank Qi Duan and Mohiuddin Ahmed for their manual assessment of the technique associations based on their expertise in the field, Vyas Sekar for providing insightful feedback on our clustering approach, and Peter Steenkiste for his valuable comments during the review of the manuscript.

Copyright 2020 Carnegie Mellon University, Eliana Christou and Rawan Al-Shaer.
This material is based upon work funded and supported by the Department of Defense under Contract No. FA8702-15-D-0002 with Carnegie Mellon University for the operation of the Software Engineering Institute, a federally funded research and development center.
References herein to any specific commercial product, process, or service by trade name, trade mark, manufacturer, or otherwise, does not necessarily constitute or imply its endorsement, recommendation, or favoring by Carnegie Mellon University or its Software Engineering Institute.

[DISTRIBUTION STATEMENT A] This material has been approved for public release and unlimited distribution.  Please see Copyright notice for non-US Government use and distribution.

Carnegie Mellon® and CERT® are registered in the U.S. Patent and Trademark Office by Carnegie Mellon University.
This material is licensed under Creative Commons Attribution-Non-Commercial-ShareAlike 4.0 International (CC BY-NC-SA 4.0) - \url{https://creativecommons.org/licenses/by-nc-sa/4.0/}.
Requests for other permissions should be directed to the Software Engineering Institute at permission@sei.cmu.edu.
DM20-0155

\printbibliography 

\begin{figure*}[h]
  \includegraphics[scale=0.25, trim={0 0 0 1cm}, clip]{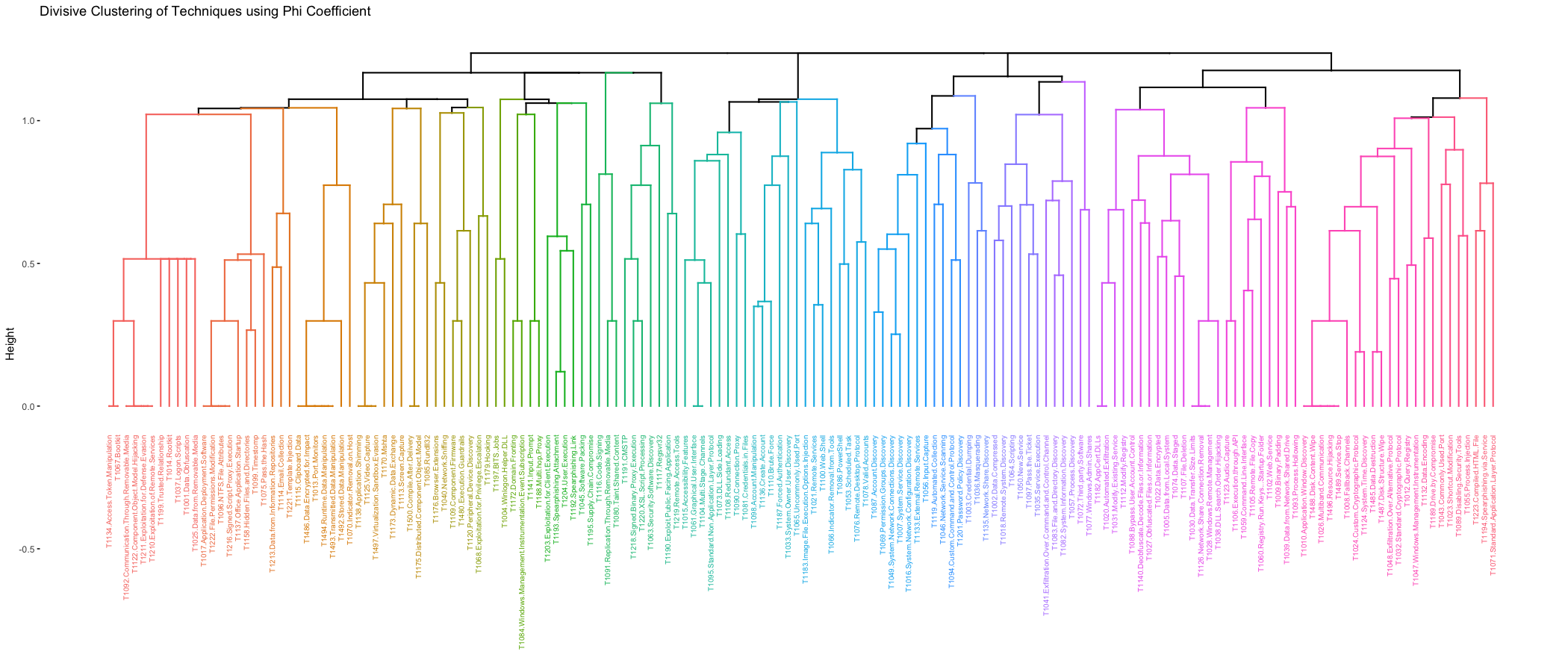}
  \caption{Divisive Hierarchical Clustering of APT Attacks}
  \label{fig:DivAPT}
\end{figure*}

\begin{figure*}[h]
  \includegraphics[scale=0.25, trim={0 0 0 1cm}, clip]{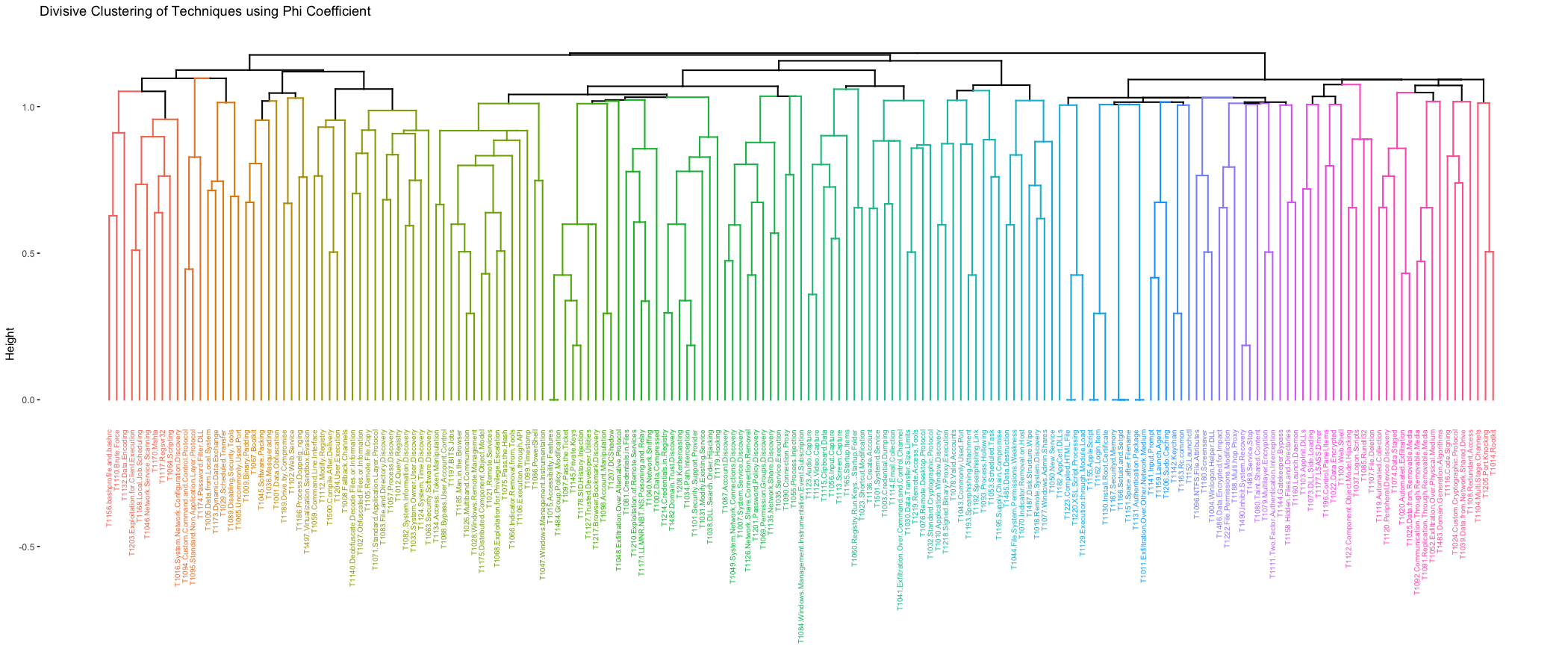}
  \caption{Divisive Hierarchical Clustering of Software Attacks}
  \label{fig:DivSoftware}
\end{figure*}

\begin{figure*}
    \centering
  \includegraphics[angle=90, scale=0.18]{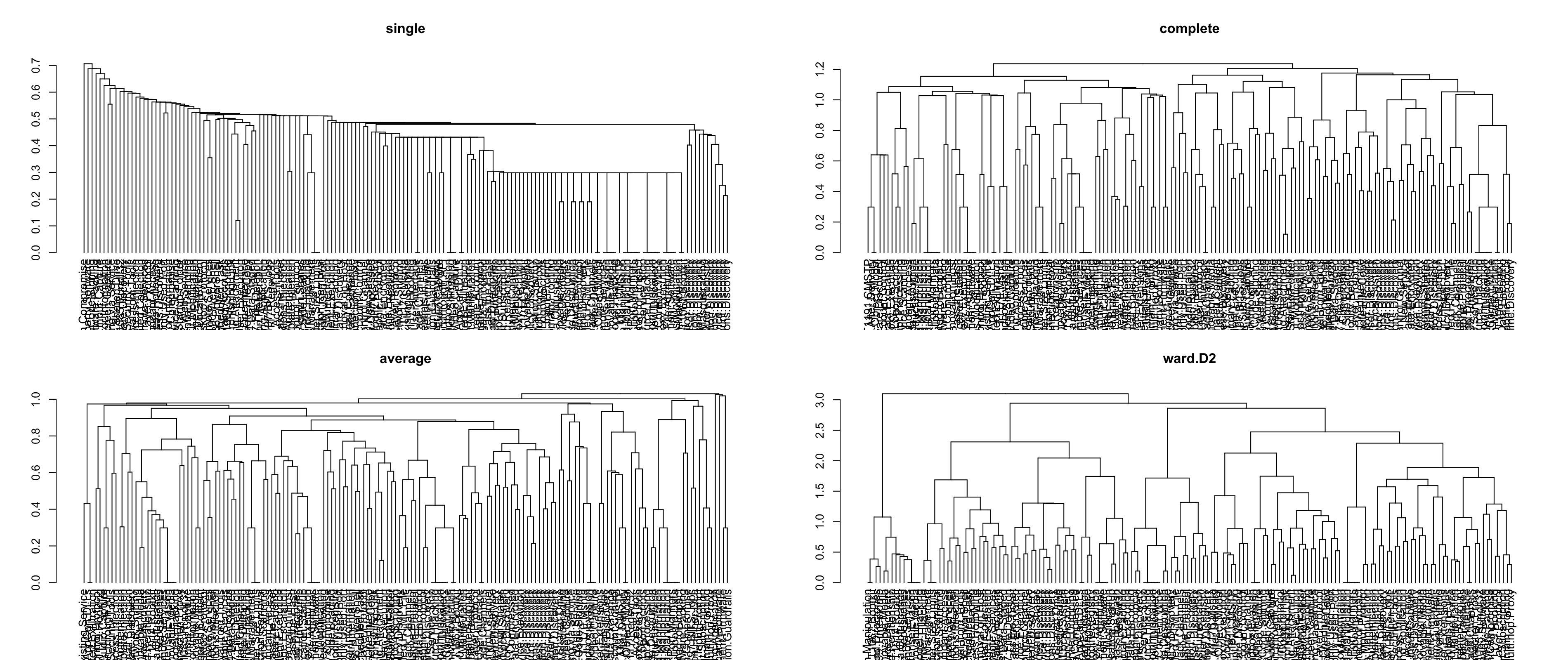}
  \caption{Agglomerative Hierarchical Clustering Linkage Methods of APT Attacks}
  \label{fig:ensemble}
\end{figure*}

\appendix
Figure~\ref{fig:DivAPT} and Figure~\ref{fig:DivSoftware} display divisive clustering results. 
Figure~\ref{fig:ensemble} displays agglomerative clustering results. 

%

\end{document}